\documentclass[11pt,reqno]{amsart}

\usepackage{amssymb,mathtools}
\usepackage{url}
\usepackage{hyperref}
\usepackage[dvips]{color}
\usepackage{tikz-cd}
\usepackage[all]{xy}
\SelectTips{cm}{} 

\allowdisplaybreaks[4]

\usepackage{mathrsfs}
\let\mathcal\mathscr
\usepackage[mathscr]{eucal}

\newcommand*{\pd}[2]{\mathchoice{\frac{\partial#1}{\partial#2}}
  {\partial#1/\partial#2}{\partial#1/\partial#2}
  {\partial#1/\partial#2}}

\let\phi=\varphi
\let\kappa=\varkappa
\let\epsilon=\varepsilon

\DeclareMathOperator{\sym}{sym}

\newcommand*{\jac}[2]{\left\{ #1,#2  \right\}}

\newcommand\restr[2]{{% we make the whole thing an ordinary symbol
  \left.\kern-\nulldelimiterspace % automatically resize the bar with \right
  #1 % the function
  \littletaller % pretend it's a little taller at normal size
  \right|_{#2} % this is the delimiter
  }}

\newcommand{\littletaller}{\mathchoice{\vphantom{\big|}}{}{}{}}

\newdir{ >}{{}*!/-5pt/\dir{>}}

\theoremstyle{theorem}
\newtheorem{proposition}{Proposition}
\newtheorem{corollary}{Corollary}
\theoremstyle{definition}
\newtheorem{example}{Example}

\theoremstyle{remark}
\newtheorem{remark}{Remark}

\usepackage{mathrsfs}
\let\mathcal\mathscr
\usepackage[mathscr]{eucal}
\newcommand{\cprime}{\/{\mathsurround=0pt$'$}}

\author{Ji\v rina Jahnov\'a, Petr~Voj{\v{c}}{\'{a}}k}
\address{Mathematical Institute, Silesian University in Opava, Na Rybn\'{\i}\v{c}ku 1, 746 01 Opava, Czech Republic}
\email{Jirina.Jahnova@math.slu.cz, Petr.Vojcak@math.slu.cz} 
\title[On recursion operators for full-fledged nonlocal symmetries of the rYME]{On recursion operators for full-fledged nonlocal symmetries of the reduced quasi-classical self-dual Yang-Mills equation}

\begin{document}

\begin{abstract}
We introduce the idea of constructing recursion operators for full-fledged nonlocal symmetries and apply it to the reduced quasi-classical self-dual Yang-Mills equation. It turns out that the discovered recursion operators can be interpreted as infinite-dimensional matrices of differential functions which act on the generating vector-functions of the nonlocal symmetries simply by matrix multiplication. We investigate their algebraic properties and discuss the $\mathbb{R}$-algebra structure on the set of all recursion operators for full-fledged nonlocal symmetries of the equation in question. Finally, we illustrate the actions of the obtained recursion operators on particularly chosen full-fledged symmetries and emphasize their advantages compared to the actions of traditionally used recursion operators for shadows. 
\end{abstract}

\subjclass[2020]{35B06}
\keywords{The reduced quasi-classical self-dual Yang-Mills equation, Lax pairs, coverings, recursion operators, nonlocal symmetries.}
\maketitle

\section*{Introduction}
\label{sec:introduction}
Partial differential equations play an important role in physics and other applied sciences because they describe many natural phenomena. Therefore, it is no wonder that they have been intensively investigated over the past few decades. A very important tool in the study of systems of nonlinear partial differential equations is the symmetry analysis: having the symmetry of the system and one of its solutions, we can construct another one. A Lie subalgebra $\mathfrak g$ of the Lie algebra of all infinitesimal symmetries of the given system can be used to determine special types of solutions - the $\mathfrak g$-invariant solutions; for many nonlinear systems, these solutions are the only exact solutions that we are able to obtain (see. e.g \cite{Olver93,kra2}), thus they are very important from the mathematical as well as physical point of view. Systems that possess infinitely many generalized symmetries are often predisposed to linearization. This can be typically achieved either through a change of variables or by employing the inverse scattering method, see e.g. \cite{Ablowitz91, Dunajski10, Olver93, Wahlquist}.

A significant role in the study of symmetries of systems of partial differential equations is played by recursion operators, because they allow us to generate infinite hierarchies of symmetries. Roughly speaking, given a differential equation, if we know one of its symmetries and if we have a recursion operator at our disposal, we can apply this operator to generate a new symmetry, then we apply the recursion operator to this new symmetry and so on, arriving at the infinite hierarchy of symmetries of the equation in question.

In the context of symmetry transformations, recursion operators were first introduced by P. J. Olver in 1977 (see \cite{Olver77}) and further used and developed by other authors, see e.g \cite{Bluman89, Fokas87, Olver93} and references therein. Within this approach the recursion operators arise as pseudodifferential operators, i.e. differential operators expressed as the linear combinations of total derivative operators and their formal inverses,  the coefficients being the (matrix valued) differential functions. So that, in general, they may contain the expressions like $D_x^{-1}$, and sometimes it may happen that it is not clear how these formal operators should act on a given symmetry. 

An alternative view on the recursion operators, based on the concept of B\"acklund transformations, was proposed in \cite{Papachristou91}, and this idea was further developed and given the rigorous mathematical background by Marvan in  \cite{Marvan95}. The general idea of this geometric approach is that a recursion operator can be viewed as a B\"acklund auto-transformation for the linearized equation, i.e. the equation defining the (generating functions, or characteristics of) symmetries of the original nonlinear PDE. 

Treating a recursion operator as a B\"acklund auto-transformation, it can be applied to any symmetry, but the result may depend on new additional quantities.  More precisely, for a given partial differential system $\mathcal{E}$, it is natural to consider its differential covering $\tilde{\mathcal{E}}$, which can be roughly viewed as an overdetermined system that introduces additional dependent variables (nonlocal variables, pseudopotentials) such that $\mathcal{E}$ implies the compatibility conditions of $\tilde{\mathcal{E}}$, and subsequently to study the recursion operators which act within the set of $\tilde{\mathcal{E}}$-shadows of nonlocal symmetries of $\mathcal{E}$. The recursion operators constructed in this way may provide us with infinite hierarchies of $\tilde{\mathcal{E}}$-shadows. Nevertheless, concerning full-fledged nonlocal symmetries the things are not so simple (see e.g. \cite{boch}): not every $\tilde{\mathcal{E}}$-shadow can be lifted to a full-fledged nonlocal symmetry of $\mathcal{E}$ (i.e. a symmetry of $\tilde{\mathcal{E}}$) in the covering $\mathcal{\tilde{E}}$. Hence, the recursion operators computed in this way (and usually they \textit{are} computed in this way) do not provide us immediately with the infinite-dimensional Lie algebra of symmetries.
However, the infinite-dimensional Lie algebras of nonlocal symmetries of $\mathcal{E}$ play an important role in the theory of integrable systems and can be used for the study of the latter (see e.g. \cite{kra1}). So, it would be very useful to know a technique that would enable us to introduce a sufficiently rich differential covering and subsequently construct recursion operators that map (in this covering) full-fledged nonlocal symmetries to other ones.

In the present paper, we develop the idea of such a technique and demonstrate it on the reduced quasi-classical self-dual Yang-Mills equation (rYME)
\begin{equation}
\label{yme}
u_{yz} = u_{tx} - u_z u_{xx}+u_xu_{xz}.
\end{equation}
To the best of our knowledge, this equation was first introduced in \cite{fer} and subsequently studied e.g. in \cite{KrugMor2016, mor, kramor}; it also appears in the context of the classification of integrable four-dimensional systems associated with sixfolds in \textbf{Gr}(4,6) as the equation arising as the compatibility condition for a canonical form of general linearly degenerate systems, see  \cite{DFKN}. Our approach to the problem of construction of ``full-fledged" recursion operators consists in exploiting certain dependencies inherent in the ``common" recursion operators that act only on the shadows of nonlocal symmetries. Based on certain observations, we present an explicit construction of the recursion operator as a standard mapping on the set of full-fledged (in the given covering) nonlocal symmetries of the rYME \eqref{yme}. However, let us point out that this our demonstration example was chosen more or less randomly and in principle we could select another one. Indeed, based on specific computational experiments performed on other equations, we  are pretty sure that the stated explicit dependency between recursion operators below can also be discovered for other 4D linearly degenerate Lax integrable equations, namely for all equations listed in Tab. 2 in \cite{DFKN} (which include the rYME, as well as, for example, the 4D Mart\'{i}nez-Alonso Shabat equation, see e.g. \cite{voj} and references therein for more details), or for the modified Mart\'{i}nez-Alonso Shabat equation \cite{Mor-Ser} (cf. also \cite{Bar}). Our approach is surely also applicable to 3D equations like the (modified) Veronese web equation, the rdDym equation, the universal hierarchy equation and the Pavlov equation, see e.g. \cite{BKMV, Dun, KMV, Kru, Zakh} for more information about these equations, although in this case the set of possible recursion operators seems to be (quite logically) much poorer (and hence less interesting) then in the case of 4D equations.

The structure of this paper is  as follows: in Section \ref{sec:1}, we recall a few necessary notions and facts from the geometrical theory of PDEs, for example the notions of linearization, symmetry, differential covering, tangent equation and recursion operator.

In Section \ref{sec:2}, we proceed to the construction of the differential covering and a ``full-fledged" recursion operator for symmetries of the rYME \eqref{yme}. In order to construct a sufficiently large covering, we use techniques that have been used successfully before, e.g. in \cite{BKMV}, \cite{KMV}, \cite{KrasVoj} or \cite{voj}. They are based on the existence of the Lax representation of the equation in question which contains two non-removable parameters $\lambda$ and $\mu$: we expand the nonlocal variable into a formal Laurent series in $\lambda$ and in $\mu$, respectively, and arrive at three inequivalent infinite-dimensional coverings. Their Whitney product forms the desired covering which enables us to find an explicit dependence between the components of full-fledged nonlocal symmetries whose shadows are related by a ``common" recursion operator. The resulting ``full-fledged" recursion operator for the rYME \eqref{yme} has surprisingly nice properties: its action on the generating vector-functions of nonlocal symmetries can be viewed just as (infinite-dimensional) matrix multiplication, the corresponding matrix having only finitely many nonzero entries in each row and its entries being only nonlocal differential functions living in the given Whitney product. 

Section \ref{sec:3} is devoted to the investigation of algebraic properties of the above constructed recursion operator. It turns out that it is an automorphism of the $\mathcal{A}$-module $\mathcal{A}^{\mathbb{N}}$ (where $\mathcal{A}$ is a suitably chosen ring of nonlocal functions living in the Whitney product) whose inverse is also a ``full-fledged" recursion operator. This implies that the presented recursion operator is an isomorphism on the real vector space of full-fledged nonlocal symmetries whose generating functions lie in $\mathcal{A}^{\mathbb{N}}$. We also find another invertible ``full-fledged" recursion operator that possesses the same nice properties as the first one. We prove that all the just obtained recursion operators and their inverses pair-wise mutually commute, which implies that the set of all the compositions of these recursion operators carries a commutative group structure with respect to the composition operation. Finally, we conjecture that the two just found ``full-fledged" recursion operators together with their inverses are the generators of the whole associative algebra of all recursion operators for the rYME \eqref{yme}.

In Section 4, we present two particular examples of infinite hierarchies of full-fledged symmetries of the rYME \eqref{yme} that are generated by our recursion operators from the chosen seed symmetries. In the end, we shortly discuss the action of the full-fledged recursion operators on symmetries belonging to the families of symmetries indexed by arbitrary functions, and emphasize its advantages compared to the action of traditionally used recursion operators for shadows.

\section{Preliminaries}
\label{sec:1}
In this section, we briefly summarize some basic notions and facts from nonlocal geometry of PDEs which are crucial for the next sections of this paper. For more detailed exposition, we refer the reader to the classical book \cite{boch}, the overview paper \cite{kra1} and/or e.g. \cite{kra2,kra3}. 

Let $\pi:\mathbb{R}^m\times\mathbb{R}^n\to\mathbb{R}^n$ be a trivial bundle where $x^1,\dots,x^n$ are coordinates standing for the independent variables in $\mathbb{R}^n$ and the coordinates $u^1,\dots,u^m$ in $\mathbb{R}^m$  denote the dependent variables. Let $\pi_{\infty}:J^{\infty}(\pi)\to\mathbb{R}^n$ be the bundle of infinite jets, the coordinates in $J^{\infty}(\pi)$ being $x^i,u^j,u_{\sigma}^j$, where $i=1,\dots,n$, $j=1,\dots,m$, and $\sigma$ are symmetric multi-indices of arbitrary finite length $|\sigma|$ that consist of the integers $1,\dots,n$; the coordinate $u^j_\sigma$ stands for the derivative of the dependent variable $u^j$ with respect to the independent variables stated in the multiindex $\sigma$ here. 
The infinite jet space $J^{\infty}(\pi)$ is equipped with the \textit{total derivatives operators}
$$D_{x^i}\equiv D_i = \frac{\partial}{\partial x^i} + \sum \limits_{j,\sigma} u_{\sigma i}^j \frac{\partial}{\partial u_\sigma^j},\quad i=1,\dots,n,$$
that provide us with the so-called  \textit{Cartan distribution} $\mathcal{C}$ on $J^{\infty}(\pi)$. Below $D_{\sigma}$ denotes the composition of the total derivative operators with respect to those independent variables that are stated in the multiindex $\sigma$. 

A \textit{differential function} on $J^{\infty}(\pi)$ is a smooth function $F$ on $J^{\infty}(\pi)$ of finite \textit{order} 
$$\mathrm{ord}\ F=\mathrm{max}\left\{|\sigma|\ |\ \partial F/\partial u^j_\sigma\neq0\right\}.$$ 
The $\mathbb{R}$-algebra of differential functions is denoted by $\mathcal{F}(\pi)$. Each function $F\in\mathcal{F}(\pi)$ of order $k$ can be identified with a nonlinear differential operator $\Delta_F(f^1,\dots,f^m)=F(x,f^j,\ldots, \frac{\partial^{|\sigma|}f^j}{\partial x^{\sigma}}, \ldots )$ of order $k$, thus each system of differential functions $\{ F^{\alpha}\} \subset \mathcal{F}(\pi)$ provides us with a system of differential equations 
$$\Delta_{F^{\alpha}}(f^1,\dots,f^m)=0,$$  where $f=(f^1(x),\dots,f^j(x),\dots,f^m(x))$ are the unknown functions in the independent variables $x=(x^1,\ldots, x^n)$, and vice versa.

Let $F=(F^1,\dots,F^{r})$ be a vector differential function on $J^{\infty}(\pi)$, i.e. $F^{\alpha}\in\mathcal{F}(\pi)$ for each $\alpha=1,\dots,r$. From now on, an \textit{(infinitely prolonged) differential equation $\mathcal{E}$} is the submanifold 
\begin{equation}
\label{eqE}
\mathcal{E}\equiv \mathcal{E}_F=\left\{\theta\in J^{\infty}(\pi)\ |\ D_{\sigma}(F^{\alpha})(\theta)=0\ \forall \alpha,\sigma\right\}\subset J^{\infty}(\pi),
\end{equation}
(it corresponds to the system of differential equations given by the functions $F^{\alpha}$ considered together with all its differential consequences).The structure of the equation $\mathcal{E}$ is given by the \textit{Cartan distribution} restricted to $\mathcal{E}$ which is spanned by the $\pi_{\infty}$-horizontal vector fields 
$$\bar D_i=\restr{D_i}{\mathcal{E}},\quad i=1,\dots,n.$$

A solution of $\mathcal{E}$ is a section $s=(u^1=s^1(x),\dots,u^m=s^m(x))$ of the bundle $\pi$ such that its infinite jet $j_{\infty}(s)=(u^1_{\sigma}=\frac{\partial^{|\sigma|}s^1}{\partial x^{\sigma}},\dots,u^m_{\sigma}=\frac{\partial^{|\sigma|}s^1}{\partial x^{\sigma}})$ (here $\sigma$ runs over all multiindices) lies in $\mathcal{E}$. The solutions correspond to the maximal $n$-dimensional integral manifolds of the Cartan distribution $\mathcal{C}$.

Suppose that it is possible to choose and fix a set $\mathbb{I}$ of internal coordinates on $\mathcal{E}$ (this can be done by locally resolving the equations $F^1=F^2=\dots=F^r=0$ with respect to certain partial derivatives). Denote ${\mathcal{F}}(\mathcal{E})$ the algebra of smooth functions on $J^{\infty}(\pi)$ restricted to $\mathcal{E}$ and expressed in the internal coordinates.
A $\pi_{\infty}$-vertical vector field 
$$S=\sum_{u_\sigma^j\in\mathbb{I}}S_{\sigma}^j\frac{\partial}{\partial u^j_{\sigma}}, \quad S_{\sigma}^j\in{\mathcal{F}}(\mathcal{E})$$
 is a \textit{(higher infinitesimal) symmetry} of the equation $\mathcal E$ if it holds $[\mathcal{C},S]\subset\mathcal{C}$. It can be shown that symmetries of the equation $\mathcal E$ are exactly the evolutionary vector fields
\begin{equation}
\label{sec1:1}
\mathbf E_{\Phi} = \sum \limits_{u_\sigma^j \in\mathbb{I}} \bar D_{\sigma}(\phi_j)\frac{\partial}{\partial u_\sigma^j}, \quad \varphi_j\in\mathcal{F}(\mathcal{E}),
\end{equation}
where $\Phi=(\phi_1,\phi_2, \ldots, \varphi_m)$ satisfies the condition
\begin{equation}
\label{lin_sym_cond}
\ell_{\mathcal E} (\Phi) \equiv \restr{\ell_F}{\mathcal E} (\Phi) = 0,
\end{equation}
and
$$
\ell_F=\begin{pmatrix}
\sum_\sigma \frac{\partial F^1}{\partial u_\sigma^1}D_\sigma & \sum_\sigma \frac{\partial F^1}{\partial u_\sigma^2}D_\sigma & \ldots \\[3mm]
\sum_\sigma \frac{\partial F^2}{\partial u_\sigma^1}D_\sigma & \sum_\sigma \frac{\partial F^2}{\partial u_\sigma^2}D_\sigma & \ldots\\
\vdots & \vdots & \ddots
\end{pmatrix}
$$
denotes the \textit{operator of linearization of $F$}. The vector-valued differential function $\Phi\in\left(\mathcal{F}(\mathcal{E})\right)^m$ is called the \textit{generating function of the symmetry} in question.  

The set of all symmetries of the equation $\mathcal E$ equipped with the commutator forms a Lie algebra $\sym(\mathcal E)$. It is isomorphic to the Lie algebra of all solutions to \eqref{lin_sym_cond}, the Lie algebra structure being given by the Jacobi bracket $\jac{\cdot}{\cdot}$ defined by the formula:
\begin{equation}
\label{sec1:2}
\jac{\Phi_1}{\Phi_2} = \mathbf E_{\Phi_1}(\Phi_2)-\mathbf E_{\Phi_2}(\Phi_1).
\end{equation}
Due to this isomorphism we will not distinguish below between the symmetries and their generating functions.

A \textit{recursion operator} for symmetries of $\mathcal{E}$ (in the classical sense of Olver, see e.g. \cite{Olver93}) is an $\mathbb{R}$- linear operator $\mathcal{R}:(\mathcal{F}(\mathcal{E}))^m\to(\mathcal{F}(\mathcal{E}))^m$ with the property that whenever $\Phi$ is a generating function of a symmetry of $\mathcal{E}$, so is $\mathcal{R}(\Phi)$.

Consider the space $\tilde{\mathcal E} = \Bbb R^l \times \mathcal E$, where $1\leq l \leq \infty$ and $\mathcal E$ is determined by \eqref{eqE}. Denote the coordinates in $\Bbb R^l$ as $w^\eta$ and assume that the space $\tilde{\mathcal E}$ is endowed with the operators
\begin{equation}
\label{cd-gen}
\tilde D_i = \bar D_i + \sum \limits_\eta X_i^\eta \pd{}{w^\eta},\quad X_i^{\eta}\in\mathcal{F}(\tilde{\mathcal E}),
\end{equation}
which pair-wise commute, i.e.
\begin{equation}
\label{cov_distr}
[\tilde D_i, \tilde D_j]=\bar D_i(X_j^\eta)-\bar D_j(X_i^\eta) + \sum \limits_\nu\left( X_i^\nu \pd{X_j^\eta}{w^\nu} - X_j^\nu \pd{X_i^\eta}{w^\nu} \right) = 0
\end{equation}
for all $i,j$ and $\eta$ (in other words, the Cartan distribution $\tilde{\mathcal{C}}$ on $\tilde{\mathcal{E}}$ given by these vector fields is integrable). We say that the projection $\tau: \tilde{\mathcal E} \rightarrow \mathcal E$ to the second factor of $\tilde{\mathcal E}$ is a \textit{(differential) covering} over the equation $\mathcal E$ and the coordinates $w^\eta$ are called \textit{nonlocal coordinates} or \textit{nonlocal variables} in this covering. The covering space $\tilde{\mathcal E}$ can be equivalently regarded as a differential equation  $\tilde{\mathcal{E}}=\mathcal{E}\cap\mathcal{E}_{\tau}\subset J^{\infty}(\tilde{\pi})$, where
the equation $\mathcal{E}_{\tau}$ is given by the relations
\begin{equation}\mathcal{E}_{\tau}:\ 
\label{eq:1r}
\frac{\partial w^{\eta}}{\partial x^i}=X_i^\eta(x,u_{\sigma}^j,w^{\gamma}), \quad i=1,\ldots,n, \; \eta=1,\ldots,l,
\end{equation}
where $w^1,\dots,w^l$ are considered to be new unknown functions of the variable $x=(x^1,\dots,x^n)$. The equation $\mathcal{E}_{\tau}$ is called the \textit{covering equation} or the \textit{Lax representation} for $\mathcal{E}$. In the case when the system \eqref{eq:1r} is given in the form of two equations for each nonlocal variable, it is referred to as a \textit{Lax pair}.  The compatibility conditions for the system \eqref{eq:1r} are differential consequences of the equation $\mathcal{E}$. 

Symmetries of the equation $\tilde{\mathcal E}$ are said to be nonlocal symmetries of the equation $\mathcal E$ in this covering. Thus,  \textit{(full-fledged) nonlocal $\tau$-symmetries of $\mathcal E$} are ${\tilde\pi}^{\infty}$-vertical vector fields $X$ on $\tilde{\mathcal{E}}$ that preserve the Cartan distribution $\tilde{\mathcal{C}}$ on $\tilde{\mathcal{E}}$, i.e. $[X,\tilde{\mathcal{C}}]\subset\tilde{\mathcal{C}}$, which means that $X$ must be exactly of the form
\begin{equation}
\label{evol-non-sym}
X\equiv\mathbf{E}_{(\Phi,\Psi)}=\sum \limits_{u_\sigma^j\in\mathbb{I}} \tilde D_{\sigma}(\phi_j)\frac{\partial}{\partial u_\sigma^j} +\sum \limits_\eta \psi^\eta \pd{}{w^\eta},
\end{equation}
where $\Phi=(\phi_1, \phi_2,\ldots,\varphi_m)$, $\Psi=(\psi_1, \psi_2, \ldots,\psi_l)$ are vector-valued functions on $\tilde{\mathcal E}$ satisfying the system
\begin{align}
\label{sec1:4}
\tilde \ell_{\mathcal E}(\Phi)&=0,\\[2mm]
\label{sec1:5}
\tilde D_i(\psi^\eta) & = \tilde \ell_{X_i^\eta}(\Phi) + \sum \limits_\nu \pd{X_i^\eta}{w^\nu}\psi^\nu,
\end{align}
and ${\tilde{\ell}}_{\mathcal{E}}$ denotes the natural lift of the differential operator ${\ell}_{\mathcal{E}}$  from $\mathcal E$ to $\tilde{\mathcal E}$. Solutions of the equation \eqref{sec1:4} are called \textit{shadows of nonlocal symmetries}. Let us point out that some authors call the shadows of nonlocal symmetries shortly as nonlocal symmetries. However, whenever we talk about $\tau$-nonlocal symmetry in the following, we always mean a full-fledged $\tau$-nonlocal symmetry.

Let $\tau_1:\tilde{\mathcal{E}}_1\to\mathcal{E}$ and $\tau_2:\tilde{\mathcal{E}}_2\to\mathcal{E}$ be two coverings over $\mathcal{E}$. Denote $v^1,v^2,\ldots,v^k$ the nonlocal variables in $\tau_1$ and  $w^1,w^2,\ldots,w^l$ the nonlocal variables in $\tau_2$, $1\leq k,l \leq \infty$. Let the Cartan distributions $\tilde{\mathcal{C}}^{1}$ on $\tilde{\mathcal{E}}_1$, resp.  $\tilde{\mathcal{C}}^{2}$ on $\tilde{\mathcal{E}}_2$ be given by the vector fields $\tilde D_i^{(1)}=\bar D_i+X_i^{(1)}$, resp. $\tilde D_i^{(2)}=\bar D_i+X_i^{(2)}$, $i=1,\dots,n$, where 
$X_i^{(1)}= \sum_{\eta}{V_i^{\rho}}\frac{\partial}{\partial v^{\rho}}$ and $X_i^{(2)}= \sum_{\eta}{W_i^{\eta}}\frac{\partial}{\partial w^{\eta}}.$
 Consider the direct product $\tilde{\mathcal{E}}_1\times\tilde{\mathcal{E}}_2$ and its subset $\tilde{\mathcal{E}}_1\oplus\tilde{\mathcal{E}}_2=\left\{(y_1,y_2)\in \tilde{\mathcal{E}}_1\times\tilde{\mathcal{E}}_2 \ |\ \tau_1(y_1)=\tau_2(y_2)\right\}$ equipped with the (integrable) distribution  ${\mathcal{C}}^{\oplus}$ defined by the fields
 $\tilde D_i^{\oplus}=\tilde D_i+ X_i^{(1)}+X_i^{(2)}$. Then the projection $\tau_1\oplus\tau_2:\tilde{\mathcal{E}}_1\oplus\tilde{\mathcal{E}}_2\to\mathcal{E}, (\tau_1\oplus\tau_2)(y_1,y_2)=\tau_1(y_1)=\tau_2(y_2)$ is called the \textit{Whitney product}
 of $\tau_1$ and $\tau_2$. The corresponding covering equation is given by the relations
 \begin{align*}
 \mathcal{E}_{\tau_1\oplus\tau_2}:\frac{\partial v^{\rho}}{\partial x^i}&=V_i^\rho(x,u_{\sigma}^j,v^{\gamma_1}),\\
 \frac{\partial w^{\eta}}{\partial x^i}&=W_i^\eta(x,u_{\sigma}^j,w^{\gamma_2}),\quad i=1,\ldots,n, \quad \rho,\gamma_1=1,\dots,k \quad \eta,\gamma_2=1,\ldots,l.
\end{align*}

The tangent equation $\mathcal{TE}\subset J^{\infty}(\bar\pi)$ of the equation $\mathcal E$ is given by the system
\begin{equation*}
  \mathcal{TE}:\
  \begin{array}{r}
    F^\alpha \left( x,\ldots, u^j_{\sigma}, \ldots \right)=0,\\ [2mm]
    \ell_F(p) = 0,
  \end{array}
\end{equation*}
where $p=(p^1,p^2,\ldots,p^m)$ are new dependent variables, the projection $t:\mathcal{T}\mathcal{E}\to\mathcal{E}, (x,u_{\sigma}^j,p_{\sigma})\mapsto (x,u_{\sigma})$ is the so-called \textit{tangent covering} of $\mathcal{E}$. The sections of $\mathcal{T}\mathcal{E}$ that map the Cartan distribution on $\mathcal{E}$ to the Cartan distribution on $\mathcal{T}\mathcal{E}$ are the symmetries of $\mathcal{E}$.

A \textit{B\"{a}cklund transformation} between the equations $\mathcal{E}_1$ and $\mathcal{E}_2$ is the diagram
\begin{figure}[h]
\begin{center}
\begin{tikzcd}[row sep=1cm, column sep=10mm]
& 
\tilde{\mathcal E}
\arrow[bend left=0, "\tau_1"{description}]{dl} 
\arrow[bend left=0, "\tau_2"{description}]{dr} 
&
\\
\mathcal E_1
&
&
\mathcal E_2\\[-13mm]
\end{tikzcd}
\end{center}
\end{figure}

\noindent where $\tau_1$ and $\tau_2$ are coverings. If $\mathcal{E}_1=\mathcal{E}_2=\mathcal{E}$, then the B\"{a}cklund transformation of $\mathcal{E}$ is called the \textit{B\"{a}cklund auto-transformation}.
Any B\"{a}cklund auto-transformation of $\mathcal{TE}$ relates (shadows of) symmetries of $\mathcal{E}$ to each other, which means that it can be interpreted as a \textit{recursion operator}. We discuss this approach widely in the next Section \ref{sec:2}, where we describe our idea of constructing of recursion operators for full-fledged nonlocal symmetries.

\section{Coverings and the recursion operator for full-fledged nonlocal symmetries of the rYME}
\label{sec:2}
Consider the rYME \eqref{yme}, where $x,y,z,t$ are the independent variables and $u$ stands for the dependent variable. The manifold $\mathcal{E}$ corresponding to this equation lies in the jet space $J^{\infty}(\pi)$, where $\pi:\mathbb{R}\times\mathbb{R}^4\to\mathbb{R}^4$. Then the internal coordinates on $\mathcal{E}$ are $u_{x^it^j},u_{x^iy^kt^j},u_{x^iz^lt^j}$ where $i,j\geq 0,\ k,l>0$, and the Cartan distribution on $\mathcal{E}$ is spanned by the total derivative operators $D_x,D_y,D_z,D_t$ on $J^{\infty}(\pi)$ restricted to $\mathcal{E}$, that is
\begin{align*}
\bar D_x&=\frac{\partial}{\partial x}+\sum_{\mathclap{\substack{i,j\geq 0,\\[1mm] k,l>0}}}\left(u_{x^{i+1}t^j}\frac{\partial}{\partial u_{x^it^j}}+u_{x^{i+1}y^kt^j}\frac{\partial }{\partial u_{x^{i}y^kt^j} }+u_{x^{i+1}z^lt^j}\frac{\partial}{\partial u_{x^iz^lt^j}}\right),\\
\bar D_y&=\frac{\partial}{\partial y}+\sum_{\mathclap{\substack{i,j\geq 0,\\[1mm] k,l>0}}}\left(u_{x^{i}yt^j}\frac{\partial}{\partial u_{x^it^j}}+u_{x^{i}y^{k+1}t^j}\frac{\partial }{\partial u_{x^{i}y^kt^j} }+D_x^{i}D_{z}^{l-1}D_{t}^j(u_{tx}-u_zu_{xx}+u_xu_{xz})\frac{\partial}{\partial u_{x^iz^lt^j}}\right),\\
\bar D_z&=\frac{\partial}{\partial z}+\sum_{\mathclap{\substack{i,j\geq 0,\\[1mm] k,l>0}}}\left(u_{x^{i}zt^j}\frac{\partial}{\partial u_{x^it^j}}+D_x^iD_y^{k-1}D_t^j(u_{tx}-u_zu_{xx}+u_xu_{xz})\frac{\partial }{\partial u_{x^{i}y^kt^j} }+u_{x^iz^{l+1}t^j}\frac{\partial}{\partial u_{x^iz^lt^j}}\right),\\
\bar D_t&=\frac{\partial}{\partial t}+\sum_{\mathclap{\substack{i,j\geq 0,\\[1mm] k,l>0}}}\left(u_{x^{i}t^{j+1}}\frac{\partial}{\partial u_{x^it^j}}+u_{x^{i}y^kt^{j+1}}\frac{\partial }{\partial u_{x^{i}y^kt^j} }+u_{x^{i}z^lt^{j+1}}\frac{\partial}{\partial u_{x^iz^lt^j}}\right).
\end{align*}

As it has already been mentioned in the introductory part of this paper, the rYME \eqref{yme} was introduced for the first time in \cite{fer}, including its one-parameter isospectral Lax representation
\begin{equation}
  \label{lax1-yme}
  s_t=u_zs_x+\kappa s_z, \quad s_y=(u_x+\kappa)s_x,
\end{equation}
cf. also \cite{DFKN}. However, it turns out that the Lax pair \eqref{lax1-yme} can be easily modified by the substitution $\kappa \rightarrow -\frac{z+\mu}{t+\lambda}$ into the Lax pair which contains two parameters and explicitly depends on the independent variables $t,z$. Indeed, it is straightforward to verify that the equations
\begin{equation}
  \label{lax-yme}
  w_t=u_zw_x-\frac{z+\mu}{t+\lambda}\,w_z, \quad w_y=u_xw_x-\frac{z+\mu}{t+\lambda}\,w_x,
\end{equation}
where $\lambda,\mu \in \Bbb R$ are spectral parameters and $w$ is a nonlocal variable, also define the Lax representation of the rYME \eqref{yme}. 

\begin{remark}
\label{rm-non-var}
In fact, the ``nonlocal variables" produced by Lax pairs for equations in more than two independent variables \textit{are not} nonlocal variables in the sense of the definition stated in Sec. \ref{sec:1}. However, as we will discuss below, these ``nonlocal variables" \textit{provide us} with an infinite number of nonlocal variables for the given equation. Thus, until we analyze this problem for the case of the rYME \eqref{yme} in more details, we will call $w$ and other variables given by the Lax pairs for $\mathcal{E}$ to be just nonlocal variables.
\end{remark}

To obtain coverings of \eqref{yme}, we consider the following scheme applied before, for example, in \cite{BKMV, KMV, voj}, cf. also \cite{Pav, Man}: 
\begin{itemize}
\item[1)] The nonlocal variable $w$ given by the Lax pair \eqref{lax-yme} is expanded in a formal series in the spectral parameter $\lambda$, resp. $\mu$:
$$w=\sum \limits_{i=-\infty}^\infty \lambda^i w_i, \; \mathrm{resp.}\quad w=\sum \limits_{i=-\infty}^\infty \mu^i w_i.$$
\item[2)] By substituting this expansion into the Lax pair \eqref{lax-yme} and eliminating the parameter $\lambda$, resp. $\mu$, we get an infinite series of nonlocal quantities $w_i$, $i \in \Bbb Z$.
\item[3)] To achieve the proper definition of nonlocal variables $w_i$, we consider two reductions of the series $w_i$: (a) the negative reduction $w_i=0$ for $i>0$ and (b) the positive reduction $w_i=0$ for $i<0$.
\item[4)] In this way, we arrive at the following three inequivalent infinite-dimensional coverings $\tau^q$, $\tau^m$ and $\tau^r$ (given in the form of Lax pairs) with the nonlocal variables $q_\alpha,m_\alpha,r_\beta$ respectively, where:
\begin{align}
  \label{yme-tauq}
  \tau^q\colon\tilde{\mathcal{E}}^q\to\mathcal{E}&\quad
                \left|\begin{array}{l}
                  q_{-1} = 0,\\
                  q_{\alpha,t} = u_zq_{\alpha,x} + q_{\alpha-1,z},\\[2pt]
                  q_{\alpha,y} = u_xq_{\alpha,x} + q_{\alpha-1,x}, \quad \alpha \geq 0;
                \end{array}\right.
  \intertext{}
  \label{yme-taum}
  \tau^m\colon\tilde{\mathcal{E}}^m\to\mathcal{E}&\quad
                \left|\begin{array}{l}
                  m_{-1} = 0,\\
                  m_{\alpha,t} = u_z m_{\alpha,x}-\displaystyle \frac{z}{t}m_{\alpha,z}-\frac{1}{t}m_{\alpha-1,z},\\[3mm]
                  m_{\alpha,y} = u_x m_{\alpha,x}-\displaystyle \frac{z}{t}m_{\alpha,x}-\frac{1}{t}m_{\alpha-1,x}, \quad \alpha \geq 0;
                \end{array}\right.
\intertext{and}
 \label{yme-taur}
  \tau^r\colon\tilde{\mathcal{E}}^r\to\mathcal{E}&\quad
\left|\begin{array}{l}
                  r_{-1} = x,\quad r_0=-u\\
                  r_{\beta,x} = u_xr_{\beta-1,x} - r_{\beta-1,y},\\[2pt]
                  r_{\beta,z} = u_zr_{\beta-1,x} - r_{\beta-1,t}, \quad \beta \geq 1.
                \end{array}\right.
\end{align}
\end{itemize} 

\begin{remark}
Let us note that the coverings $\tau^q$ and $\tau^m$ are non-Abelian, whereas $\tau^r$ is Abelian. Thus the nonlocal variables $r_\beta$, $\beta\geq 1$, are interconnected with infinite series of two-component nonlocal conservation laws of the rYME \eqref{yme}. 
\end{remark}

\begin{remark} 
Whereas the coverings $\tau^q$ and $\tau^r$ can also be derived from the one-parameter Lax pair \eqref{lax1-yme}, the covering $\tau^m$ can be obtained only from the two-parameter Lax representation \eqref{lax-yme}.  
\end{remark}

As noted in Remark \ref{rm-non-var}, the equations \eqref{yme-tauq}, \eqref{yme-taum} and \eqref{yme-taur} are not in the form of the covering equation \eqref{eq:1r}. In order to  identify the nonlocal coordinates provided by these formulas, we have to introduce the infinite number of formal variables $q_{\alpha}^{a,b}$, $m_{\alpha}^{a,b}$ and $r_{\beta}^{a,b},$ $a,b,=0,1,\dots$, where $q_{\alpha}^{0,0}=q_{\alpha}$, $m_{\alpha}^{0,0}=m_{\alpha}$, $r_{\beta}^{0,0}=r_{\beta}$  and 
\begin{equation*}
q_{\alpha,x}^{a,b}=q_{\alpha}^{a+1,b}, \quad q_{\alpha,z}^{a,b}=q_{\alpha}^{a,b+1},\quad q_{\alpha,y}^{a,b}=(u_xq_{\alpha}^{1,0} + q_{\alpha-1}^{1,0})_{x^{a}z^b},\quad q_{\alpha,t}^{a,b}=(u_zq_{\alpha}^{1,0} + q_{\alpha-1}^{0,1})_{x^{a}z^b},
\end{equation*}
\vspace{1mm}
\begin{equation*}m_{\alpha,x}^{a,b}=m_{\alpha}^{a+1,b},\quad m_{\alpha,z}^{a,b}=m_{\alpha}^{a,b+1},\quad m_{\alpha,y}^{a,b}=\left(u_x m_{\alpha}^{1,0}-\displaystyle \frac{z}{t}m_{\alpha}^{1,0}-\frac{1}{t}m_{\alpha-1}^{1,0}\right)_{x^{a}z^b},\end{equation*}
\begin{equation*}m_{\alpha,t}^{a,b}=\left(u_z m_{\alpha}^{1,0}-\displaystyle \frac{z}{t}m_{\alpha}^{0,1}-\frac{1}{t}m_{\alpha-1}^{0,1}\right)_{x^{a}z^b},\end{equation*}
\vspace{1mm}
\begin{equation*}
r_{\beta,y}^{a,b}=r_{\beta}^{a+1,b},\quad r_{\beta,t}^{a,b}=r_{\beta}^{a,b+1},\quad r_{\beta,x}^{a,b}=( u_xr_{\beta-1,x} - r_{\beta-1}^{1,0})_{y^at^b},\quad r_{\beta,z}^{a,b}=( u_zr_{\beta-1,x} - r_{\beta-1}^{0,1})_{y^at^b}.
\end{equation*}
\\
Note that all the right-hand sides of the equalities above are (even though sometimes inductively) expressible in terms of the variables $q_{\alpha}^{a,b}$, $m_{\alpha}^{a,b}$ and $r_{\beta}^{a,b},$ thus, they provide us with the covering equations $\mathcal{E}_{\tau^q}$, $\mathcal{E}_{\tau^m}$ and $\mathcal{E}_{\tau^r}$ in the form \eqref{eq:1r}, $q_{\alpha}^{a,b}$, $m_{\alpha}^{a,b}$ and $r_{\beta}^{a,b},$ $\alpha=0,1,\dots,\beta=1,2,\dots$, being the true \textit{nonlocal variables}.

The total derivative operators in the covering $\tau^q$ now take the form (cf. \eqref{cd-gen})
$$\tilde{D}_x^{(q)}=\bar D_x+X^{(q)},\quad \tilde{D}_y^{(q)}=\bar D_y+Y^{(q)},\quad \tilde{D}_z^{(q)}=\bar D_z+Z^{(q)},\quad \tilde{D}_t^{(q)}=\bar D_t+T^{(q)},$$
where 
\begin{align*}
X^{(q)}&=\sum_{\alpha=1}^{\infty}\sum_{a,b=0}^{\infty}q_{\alpha}^{a+1,b}\frac{\partial}{\partial q_{\alpha}^{a,b}},\qquad &Y^{(q)}=\sum_{\alpha=1}^{\infty}\sum_{a,b=0}^{\infty}(u_xq_{\alpha,x} + q_{\alpha-1,x})_{x^{a}z^b}\frac{\partial}{\partial q_{\alpha}^{a,b}},\\ Z^{(q)}&=\sum_{\alpha=1}^{\infty}\sum_{a,b=0}^{\infty}q_{\alpha}^{a,b+1}\frac{\partial}{\partial q_{\alpha}^{a,b}},
&T^{(q)}=\sum_{\alpha=1}^{\infty}\sum_{a,b=0}^{\infty}(u_zq_{\alpha,x} + q_{\alpha-1,z})_{x^{a}z^b}\frac{\partial}{\partial q_{\alpha}^{a,b}}.
\end{align*}
The total derivative operators $D_x^{(m)}$, $D_x^{(r)}$, $D_y^{(m)}$, $D_y^{(r)}$, etc. in the coverings $\tau^m$ and $\tau^r$ are given in a similarly way.

In what follows, we will consider the Whitney product $\tau^W=\tau^q \oplus \tau^m \oplus \tau^r$ of all three coverings and carry out all calculations in $\tau^W$. The covering space $\tilde{\mathcal E}^W=\mathcal{E}_{\tau^q \oplus \tau^m \oplus \tau^r}\cap\mathcal{E}$ can be considered to be an equation in the jet space $J^{\infty}(\tilde\pi)$  given by the rYME \eqref{yme} and Eqs. \eqref{yme-tauq}--\eqref{yme-taur} for $\alpha \geq 0$, $\beta \geq 1$, the coordinates in $J^{\infty}(\tilde\pi)$ being $x,y,z,t,u_{\sigma},q_{\alpha,\sigma}^{a,b},m_{\alpha,\sigma}^{a,b},r_{\beta,\sigma}^{a,b},\  a,b\geq 0$. The Cartan distribution on $\tilde{\mathcal{E}}^W$ is spanned by the total derivative operators
 \begin{align*}
 \tilde D_x&=\bar D_x+X^{(q)}+X^{(m)}+X^{(r)}, & \tilde D_y&=\bar D_y+Y^{(q)}+Y^{(m)}+Y^{(r)},\\ 
 \tilde D_z&=\bar D_z+Z^{(q)}+Z^{(m)}+Z^{(r)},  & \tilde D_t&=\bar D_t+T^{(q)}+T^{(m)}+T^{(r)}.
 \end{align*}
 
As follows directly from \eqref{evol-non-sym}-\eqref{sec1:5}, every full-fledged nonlocal symmetry of $\mathcal{E}$ in the Whitney product $\tau^W$ (for shortness we will write only $\tau^W$-symmetry below) is of the form
$$\mathbf{E}_P=\mathbf{E}^W_{p_0}+\sum_{\alpha=0}^{\infty}\sum_{a,b=0}^{\infty}\left(\tilde D_x^a\tilde D_z^b(p_{\alpha}^{q})\frac{\partial}{\partial q_{\alpha}^{a,b}}+\tilde D_x^a\tilde D_z^b(p_{\alpha}^{m})\frac{\partial}{\partial m_{\alpha}^{a,b}}+\tilde D_y^a\tilde D_t^b(p_{\alpha+1}^{r})\frac{\partial}{\partial r_{\alpha+1}^{a,b}}\right),$$
where
$$\mathbf{E}^W_{p_0}=\sum_{\mathclap{\substack{i,j\geq 0,\\[1mm] k,l>0}}}\left(\tilde D_x^i\tilde D_z(p_0)\frac{\partial}{u_{x^it^j}}+\tilde D_x^i\tilde D_y^k\tilde D_z^j(p_0)\frac{\partial}{\partial u_{x^iy^kt^j}}+\tilde D_x^i\tilde D_z^l\tilde D_t^j(p_0)\frac{\partial}{\partial u_{x^iz^lt^j}}\right),$$
and  $p_0, p_{\alpha}^{q},p_{\alpha}^{m}, p_1^{r}, p_{\beta}^{r}$, $\alpha\geq 0,\beta\geq 2$, are smooth functions on $\tilde{\mathcal{E}}^W$ satisfying the conditions:
\begin{align}
\label{nsym-yme1}
\tilde D_{yz}(p_0) - \tilde D_{tx}(p_0) + u_z \tilde D_{xx}(p_0)-u_x\tilde D_{xz}(p_0)+u_{xx}\tilde D_z(p_0)-u_{xz}\tilde D_x(p_0)&=0,\tag{NS1}\\[2mm]
\label{nsym-yme2}
\tilde D_t(p_\alpha^q) - u_z\tilde D_x(p_\alpha^q) - q_{\alpha,x}\tilde D_z(p_0)-\tilde D_z(p_{\alpha-1}^q)&=0\tag{NS2},\\[2pt]
\label{nsym-yme3}
\tilde D_y(p_\alpha^q) -u_x\tilde D_x(p_\alpha^q) - q_{\alpha,x}\tilde D_x(p_0)-\tilde D_x(p_{\alpha-1}^q)&=0\tag{NS3},\\[2pt]
\label{nsym-yme4}
\tilde D_t(p_\alpha^m) -u_z\tilde D_x(p_\alpha^m) - m_{\alpha,x}\tilde D_z(p_0)+\frac{z}{t}\tilde D_z(p_\alpha^m)+\frac{1}{t}\tilde D_z(p_{\alpha-1}^m)&=0\tag{NS4},\\[2mm]
\label{nsym-yme5}
\tilde D_y(p_\alpha^m) -u_x\tilde D_x(p_\alpha^m) - m_{\alpha,x}\tilde D_x(p_0)+\frac{z}{t}\tilde D_x(p_\alpha^m)+\frac{1}{t}\tilde D_x(p_{\alpha-1}^m)&=0\tag{NS5},\\[2mm]
\label{nsym-yme6}
\tilde D_x(p_1^r) - \tilde D_y(p_0)+2u_x\tilde D_x(p_0)&=0\tag{NS6},\\[2pt]
\label{nsym-yme7}
\tilde D_z(p_1^r) -\tilde D_t(p_0)+u_x\tilde D_z(p_0)+u_z\tilde D_x(p_0)&=0\tag{NS7},\\[2pt]
\label{nsym-yme8}
\tilde D_x(p_\beta^r) - u_x\tilde D_x(p_{\beta-1}^r)-r_{\beta-1,x}\tilde D_x(p_0)+D_y(p_{\beta-1}^r)&=0\tag{NS8},\\[2pt]
\label{nsym-yme9}
\tilde D_z(p_\beta^r) - u_z\tilde D_x(p_{\beta-1}^r)-r_{\beta-1,x}\tilde D_z(p_0)+D_t(p_{\beta-1}^r)&=0,\tag{NS9}
\end{align}
where $p_{-1}^q=p_{-1}^m=0$.

Thus, any $\tau^W$-symmetry can be (uniquely) represented by the vector-valued generating function $P=\left[p_0, p_0^q, p_0^m, p_1^r,p_1^q, p_1^m, p_2^r,p_2^q, p_2^m\ldots  \right]$. From now on, unless otherwise stated, by \textit{a nonlocal symmetry of $\mathcal{E}$} we mean a generating function of a $\tau^W$-symmetry of $\mathcal{E}$, and  $\mathrm{sym}^{\tau^W}(\mathcal{E})$ will denote the set of all generating functions of $\tau^W$-symmetries of $\mathcal{E}$ (equipped with the Lie algebra structure given by the Jacobi bracket). 
\begin{remark}
Let us note that we have used the formal nonlocal variables $q_{\alpha}^{a,b},m_{\alpha}^{a,b},r_{\beta}^{a,b}$ in order to justify the general form of the (generating function of) $\tau^W$-symmetry of $\mathcal{E}$. However, since the equalities $q_{\alpha,x^az^b}\equiv q_{\alpha,x^az^b}^{0,0}=q_{\alpha}^{a,b},m_{\alpha,x^az^b}\equiv m_{\alpha,x^az^b}^{0,0}=m_{\alpha}^{a,b} ,r_{\beta,y^at^b}\equiv r_{\beta,y^at^b}^{0,0}=r_{\beta}^{a,b} $ hold on $\tilde{\mathcal{E}}^{W}$, we will use the more common notation $q_{\alpha,x^az^b},m_{\alpha,x^az^b},r_{\beta,y^at^b}$ for them. Let us also emphasize that we sometimes use terms like $m_{\alpha,t}, r_{\beta,x}$, etc., in the explicit formulas presented below, even though $m_{\alpha,t}, r_{\beta,x}$, etc., are not internal coordinate on $\tilde{\mathcal{E}}^W.$ In fact, such terms are used to rewrite the huge resulting formulas into a more concise form.
\end{remark}

The task to find a recursion operator for  $\tau^W$-symmetries for $\mathcal{E}$ now coincides with the problem to find a B\"{a}cklund auto-transformation of the tangent equation  $\mathcal T \tilde{\mathcal E}^W$ given by the equations \eqref{yme} and \eqref{nsym-yme1}-\eqref{nsym-yme9} with the new dependent variables $p_{0,\sigma},p_{\alpha,x^a z^b}^q, p_{\alpha,x^a z^b}^m, p_{\beta,y^a t^b}^r$, the tangent covering being
\begin{align*} 
t^W: \mathcal T \tilde{\mathcal E}^W \rightarrow \tilde{\mathcal E}^W,\quad (x,y,z,t,u_\sigma^j,q_{\alpha,x^a z^b},m_{\alpha,x^a z^b},&r_{\beta,y^a t^b}, p_{0,\sigma},p_{\alpha,x^a z^b}^q, p_{\alpha,x^a z^b}^m, p_{\beta,y^a t^b}^r)\\
& \mapsto (x,y,z,t,u_\sigma^j,q_{\alpha,x^a z^b},m_{\alpha,x^a z^b},r_{\beta,y^a t^b}).
\end{align*}
The most straightforward way is to look for a B\"{a}cklund auto-transformation expressible in the terms of the existing variables, i.e. to consider the covering equation common for both the copies of $\mathcal{T}{\tilde{\mathcal{E}}}^W$ to be just  $\mathcal{T}{\tilde{\mathcal{E}}}^W$ itself.
Due to the $\mathbb{R}$-linearity requirement imposed on the sought operator, we look for the desired relationship between the solutions to $\mathcal{T}{\tilde{\mathcal{E}}}^W$ in the following form: the components of the `new' solution to $\mathcal{T}{\tilde{\mathcal{E}}}^W$ are to be (finite) linear combinations of the components $p_{0,\sigma},p_{\alpha,x^a z^b}^q, p_{\alpha,x^a z^b}^m, p_{\beta,y^a t^b}^r$ of the `old' solution  to $\mathcal{T}{\tilde{\mathcal{E}}}^W$,  the coefficients being arbitrary functions of the internal coordinates on ${\tilde{\mathcal E}}^W$. 
This `symmetry', if it exists and we are able to describe it in an usable form, provides us with a recursion operator for $\tau^W$-symmetries of $\mathcal E$. 

The issue described above can be executed in the following way:
we suppose that the quantity $P=\left[p_0, p_0^q, p_0^m, p_1^r,p_1^q, p_1^m, p_2^r,p_2^q, p_2^m,\ldots  \right]$ is a $\tau^W$-symmetry of the rYME \eqref{yme}, i.e. the components of $P$ satisfy the {conditions} \eqref{nsym-yme1}-\eqref{nsym-yme9} on {${\tilde{\mathcal{E}}}^W$}, and subsequently try to solve Eqs. \eqref{nsym-yme1}-\eqref{nsym-yme9} {with respect to the unknown} quantity $\hat P=\left[\hat p_0, \hat p_0^q, \hat p_0^m, \hat p_1^r,\hat p_1^q, \hat p_1^m, \hat p_2^r, \hat p_2^q, \hat p_2^m,\ldots  \right]$ under the assumption that the components of $\hat P$ are linear combinations of the components of $P$ and their derivatives up to some finite order, the coefficients being (possibly) arbitrary functions of the internal coordinates on $\tilde{\mathcal E}^W$. 

Note that even though the idea of how to find a full-fledged recursion operator is quite straightforward, the implementation of the latter is not simple at all. The main difficulty of this construction consists in the fact that the recursion operator we are looking for is to act on an infinite sequence resulting in another infinite sequence, {so that} our success depends on whether we are able to find general formulas describing the relationships between all the components of the pre-image and its image. 

One but not the only one result (see Sec. \ref{sec:3} below) we obtained within this approach is the following:
\begin{proposition}
\label{yme-ro}
Let $P=\left[p_0, p_0^q, p_0^m, p_1^r,p_1^q, p_1^m, p_2^r,p_2^q, p_2^m,\ldots  \right]$ be a $\tau^W$-symmetry of the rYME \eqref{yme}, and let $\hat P=\left[\hat p_0, \hat p_0^q, \hat p_0^m, \hat p_1^r,\hat p_1^q, \hat p_1^m, \hat p_2^r, \hat p_2^q, \hat p_2^m,\ldots  \right]$ be a vector-valued function such that for each $\alpha\geq 0, \beta \geq 1$ it holds
\begin{align}
\label{yme-ro1-1}
\hat p_0 &= u_x p_0+p_1^r,\\[1mm]
\label{yme-ro1-2}
\hat p_\alpha^q &= q_{\alpha,x} p_0 + p_{\alpha-1}^q,\\[1mm]
\label{yme-ro1-3}
\hat p_\alpha^m &= m_{\alpha,x} p_0 -\frac{1}{t} p_{\alpha-1}^m - \frac{z}{t}  p_\alpha^m ,\\[1mm]
\label{yme-ro1-4}
\hat p_\beta^r &= r_{\beta,x} p_0 - p_{\beta+1}^r.
\end{align}
Then $\hat P$ is also a $\tau^W$-symmetry of the rYME \eqref{yme}.
\end{proposition}
\begin{proof}
The proof is done by straightforward computations and is presented in Appendix A.
\end{proof}

\begin{remark} 
According to Prop. \ref{yme-ro}, Eqs. \eqref{yme-ro1-1}--\eqref{yme-ro1-4} provide us with a relation which maps any known $\tau^W$-symmetry $P$ to a new $\tau^W$-symmetry $\hat P$. Hence, they can be viewed as the determining equations of a recursion operator for $\tau^W$-symmetries of the rYME \eqref{yme}. In what follows, we will denote this recursion operator as $\mathcal{R}^q$. The notation will be clarified a bit later in Remark \ref{RO-notation}, see Sec. \ref{sec:3} below.
\end{remark}

\begin{remark}
\label{ROq-mat}
One can also easily observe that Eqs. \eqref{yme-ro1-1}--\eqref{yme-ro1-4} can be rewritten in the form 
$${\hat P}^T=\mathcal{R}^q\cdot P^{T},$$
where ${}^T$ denotes the transpose of the row vectors $P$ and $\hat P$, respectively, $\cdot$ is the usual matrix multiplication, and the recursion operator $\mathcal{R}^q$ is considered to be an infinite-dimensional matrix whose structure is illustrated by its $13 \times 16$ left-upper corner as follows:
\begin{equation*}
{\mathcal R^q}=
\begin{pmatrix}
u_x & 0 & 0 & 1 & 0 & 0 & 0 & 0 & 0 & 0 & 0 & 0 & 0 & 0 & 0 & 0 &\ldots\\[2mm]
q_{0,x} & 0 & 0 & 0 & 0 & 0 & 0 & 0 & 0 & 0 & 0 & 0 & 0 & 0 & 0 & 0 &\ldots\\[2mm]
m_{0,x} & 0 & -z/t & 0 & 0 & 0 & 0 & 0 & 0 & 0 & 0 & 0 & 0 & 0 & 0 & 0 &\ldots\\[2mm]
r_{1,x} & 0 & 0 & 0 & 0 & 0 & -1 & 0 & 0 & 0 & 0 & 0 & 0 & 0 & 0 & 0 &\ldots\\[2mm]
q_{1,x} & 1 & 0 & 0 & 0 & 0 & 0 & 0 & 0 & 0 & 0 & 0 & 0 & 0 &0 & 0 &\ldots\\[2mm]
m_{1,x} & 0 & -1/t & 0 & 0 & -z/t & 0 & 0 & 0 & 0 & 0 & 0 & 0 & 0 & 0 & 0 &\ldots\\[2mm]
r_{2,x} & 0 & 0 & 0 & 0 & 0 & 0 & 0 & 0 & -1 & 0 & 0 & 0 & 0 & 0 & 0 &\ldots\\[2mm]
q_{2,x} & 0 & 0 & 0 & 1 & 0 & 0 & 0 & 0 & 0 & 0 & 0 & 0 & 0 & 0 & 0 &\ldots\\[2mm]
m_{2,x} & 0 & 0 & 0 & 0 & -1/t & 0 & 0 & -z/t & 0 & 0 & 0 & 0 & 0 & 0 & 0 &\ldots\\[2mm]
r_{3,x} & 0 & 0 & 0 & 0 & 0 & 0 & 0 & 0 & 0 & 0 & 0 & -1 & 0 & 0 & 0 &\ldots \\[2mm]
q_{3,x} & 0 & 0 & 0 & 0 & 0 & 0 & 1 & 0 & 0 & 0 & 0 & 0 & 0 & 0 & 0 &\ldots\\[2mm]
m_{3,x} & 0 & 0 & 0 & 0 & 0 & 0 & 0 & -1/t & 0 & 0 & -z/t & 0 & 0 & 0 & 0 &\ldots\\[2mm]
r_{4,x} & 0 & 0 & 0 & 0 & 0 & 0 & 0 & 0 & 0 & 0 & 0 & 0 & 0 & 0 & -1 &\ldots\\[2mm]
\vdots & \vdots & \vdots & \vdots & \vdots & \vdots & \vdots & \vdots & \vdots & \vdots & \vdots & \vdots & \vdots & \vdots & \vdots & \vdots & \ddots
\end{pmatrix}
\end{equation*}

As it can be easily verified, each column of the matrix $\mathcal R^q$ is itself a $\tau^W$-symmetry of the rYME \eqref{yme}. However, this is perfectly natural, because if the matrix $\mathcal R^q$ is supposed to represent a recursion operator, then the images of the simplest nontrivial $\tau^W$-symmetries $(1,0,0,\ldots), (0,1,0,\ldots),$ etc. must again be $\tau^W$-symmetries.
\end{remark}

\begin{remark}
\label{ROq_shad}
Let us also note that using Eqs. \eqref{nsym-yme6}--\eqref{nsym-yme7} we can readily eliminate the quantity $p_1^r$ from Eq. \eqref{yme-ro1-1}. In this way, we obtain the conventional form (i.e. the form most frequently used in the current literature, cf. e.g. \cite{Bar, BKMV, KMV, Serg, voj}) of the recursion operator $\mathcal R^q_{sh}$ for shadows of $\tau^W$-symmetries of the rYME \eqref{yme}, which reads
\begin{equation}
\begin{aligned}
\label{ro1-shad}
\mathcal R^q_{sh}: \hat p_{0,x} &= p_{0,y}-u_x p_{0,x} + u_{xx} p_0,\\
\hat p_{0,z} &= p_{0,t}-u_z p_{0,x} + u_{xz} p_0.
\end{aligned}
\end{equation}

As far as we know, even this recursion operator for shadows has not been, quite surprisingly, presented in literature yet. 
\end{remark}

\section{Properties of the recursion operator $\mathcal{R}^q$ and other recursion operators for $\tau^W$-symmetries of the rYME} 
\label{sec:3}

In order to study the recursion operator $\mathcal{R}^q$ in more detail, it is necessary to determine the $\mathbb{R}$-algebra $\mathcal{A}=\mathcal{A}(\tilde{\mathcal{E}}^W)$ of  functions (of the internal variables) on $\tilde{\mathcal{E}}^W$ that we take {into} consideration as the components of the generating functions of the $\tau^W$-symmetries of $\tilde{\mathcal{E}}^W$.

According to the general theory (see e.g. \cite{boch}), working in the coordinates in $J^{\infty}(\tilde\pi)$, the components of the generating functions of $\tau^W$-symmetries are required to arise as the restrictions of differential functions on $J^{\infty}(\tilde\pi)$ to $\tilde{\mathcal{E}}^W$. However, even functions smooth on $J^{\infty}(\tilde{\pi})$ expressed in the internal coordinates on $\tilde{\mathcal{E}}^W$ may not (as the functions of the internal coordinates) remain smooth globally any more. Thus, working in the internal coordinates on $\tilde{\mathcal{E}}^W$, it is reasonable to require the functions from $\mathcal{A}$ to be smooth only locally. For our purposes, it will be convenient to consider $\mathcal{A}$ to be the $\mathbb{R}$-algebra generated by the union $\mathcal{L}\cup\mathcal{F}$, where $\mathcal{F}$ denotes the algebra of smooth functions of the internal coordinates on  $\tilde{\mathcal{E}}^W$ and $\mathcal{L}$ is the algebra of all Laurent polynomial functions \footnote{A \textit{Laurent polynomial in the indeterminates} $X_1,\dots X_s$ with real coefficients is a polynomial in the indeterminates $X_1,X_1^{-1},\dots,X_s,X_s^{-1}$ with real coefficients}  of the internal variables that depend on only finitely many of these variables
(we will refer to the just defined functions as the \textit{differential Laurent polynomial functions on $\tilde{\mathcal{E}}^W$} below).
Note that in case we are interested just in polynomial symmetries we can put $\mathcal{L}$ instead of $\mathcal{A}$ everywehere below and all the results will still remain valid.

 Let $\mathcal{A}^{\mathbb{N}}=({\mathcal{A}(\tilde{\mathcal{E}}^W)})^{\mathbb{N}}$ denote the real vector space of all sequences of admissible functions on the equation $\tilde{\mathcal{E}}^W$, let ${\mathrm{sym}}_{\mathcal{A}}^{\tau^W}(\mathcal{E})={\mathrm{sym}}_{\mathcal{A}}(\tilde{\mathcal{E}}^W)\subset\mathcal{A}^{\mathbb{N}}$ denote the Lie algebra of the generating functions of full-fledged nonlocal $\tau^W$-symmetries for \eqref{yme} whose components  lie in ${{\mathcal{A}}}$. 
% (obviously, ${\mathrm{sym}}_{\mathcal{L}}^{\tau^W}(\mathcal{E})$ is a vector subspace of $\mathcal{L}^{\mathbb{N}}$, and in fact, it carries also the Lie algebra structure with respect to the Jacobi bracket defined in section \ref{sec:1} ). 
 The mapping $\mathcal{R}^q$ introduced in Sec. \ref{sec:2} can be readily seen to be a mapping  $\mathcal{A}^{\mathbb{N}}\to\mathcal{A}^{\mathbb{N}}$ that is $\mathbb{R}$-linear and  leaves the vector subspace $\mathrm{sym}_{\mathcal{A}}^{\tau^W}(\mathcal{E})$ of $\mathcal{A}^{\mathbb{N}}$ invariant, i.e. ${{\mathcal{R}}}^q(\mathrm{sym}_{\mathcal{A}}^{\tau^W}(\mathcal{E}))\subset\mathrm{sym}_{\mathcal{A}}^{\tau^W}(\mathcal{E})$. Thus, it is a \textit{recursion operator} in the sense of the standard definition presented in Sec. \ref{sec:1}. Moreover, since ${{\mathcal{R}}}^q$ acts on $\mathcal{A}^{\mathbb{N}}$ just as the matrix multiplication operator described above, it can be easily concluded that ${\mathcal{R}}^q$ is also an $\mathcal{A}$-linear mapping:

\begin{corollary}
The recursion operator $\mathcal{R}^q$ is an $\mathcal{A}$-module  ${\mathcal{A}}^{\mathbb{N}}$ endomorphism  that leaves the set $\mathrm{sym}_{\mathcal{A}}^{\tau^W}(\mathcal{E})$ invariant.
\end{corollary}

It turns out that there exists another recursion operator for $\tau^W$-symmetries of the rYME \eqref{yme} within the class of $\mathcal{A}$-linear mappings.
\begin{proposition}
\label{ROm}
Let $\mathcal{I}$ be the identity mapping on $\mathcal{A}^{\mathbb{N}}$. Then the $\mathcal{A}$-linear mapping $\mathcal{R}^m:\mathcal{A}^{\mathbb{N}}\to\mathcal{A}^{\mathbb{N}}$ defined by the formula
\begin{equation}
\label{yme-ro2}
\mathcal R^m = t\, \mathcal R^q + z\, \mathcal I,
\end{equation}
is a recursion operator for $\tau^W$-symmetries of the rYME \eqref{yme}.
\end{proposition}

\begin{proof}
The proof is done by straightforward computations and is presented in Appendix B.
\end{proof}
 
\begin{remark}
Note that the recursion operator $\mathcal{R}^m$ can be viewed as a matrix operator too, since the identity mapping $\mathcal{I}$ can be represented by the infinite-dimensional unit matrix. The $10 \times 14$ left-upper corner of the corresponding matrix is presented in Appendix C. Similarly as in the case of $\mathcal{R}^q$ (see Remark \ref{ROq-mat}) we can observe that each column of the matrix $\mathcal{R}^m$ is itself a  $\tau^W$-symmetry of the rYME \eqref{yme}.
\end{remark}

\begin{remark}
As it can be seen immediately from the definition of the operator $\mathcal{R}^m$ (c.f.\eqref{yme-ro1-1} and \eqref{yme-ro2}), the shadow $p_0$ of a symmetry $P \in {\mathrm{sym}}_{\mathcal{A}}^{\tau^W}(\mathcal{E})$ and the shadow $\bar p_0$ of its $\mathcal{R}^m$-image $\bar P = \mathcal R^m(P)$ are related by the formula
\begin{equation}
\label{ROm-shad}
\bar p_0 = (tu_x + z) p_0 + t p_1^r.
\end{equation}
As in the case of the recursion operator $\mathcal R^q$ (see Remark \ref{ROq_shad}), we can eliminate the quantity $p_1^r$ from \eqref{ROm-shad}, which will provide us with the conventional form of the recursion operator $\mathcal R^m_{{\mathrm sh}}$ for shadows of $\tau^W$-symmetries of the rYME \eqref{yme}:
\begin{equation}
\begin{aligned}
\label{ro2-shad}
\mathcal R^m_{sh}: \bar p_{0,x} &= t(p_{0,y}-u_x p_{0,x} + u_{xx} p_0)+zp_{0,x},\\
\bar p_{0,z} &= t(p_{0,t}-u_z p_{0,x} + u_{xz} p_0)+zp_{0,z}+p_0.
\end{aligned}
\end{equation}
\end{remark}

\begin{proposition}\label{invRO10}
The $\mathcal{A}$-modul endomorphisms ${{\mathcal{R}}}^q:{\mathcal{A}}^{\mathbb{N}}\to {\mathcal{A}}^{\mathbb{N}}$, resp. ${{\mathcal{R}}}^m:{\mathcal{A}}^{\mathbb{N}}\to {\mathcal{A}}^{\mathbb{N}}$ are bijective, and their inverses $({{\mathcal{R}}}^{q})^{-1}$, resp. $({{\mathcal{R}}}^{m})^{-1}$ are given by the formulas
\begin{align}
\label{yme-ro1-inv1}
p_0 &= \frac{\hat p_0^q}{q_{0,x}},\\[1mm]
\label{yme-ro1-inv2}
p_\alpha^q &= -\frac{q_{\alpha+1,x}}{q_{0,x}} \hat p_0^q + \hat p_{\alpha+1}^q,\\[1mm]
\label{yme-ro1-inv3}
p_\alpha^m &=  \sum \limits_{j=0}^\alpha \frac{t\hat p_j^m}{(-z)^{\alpha-j+1}}-\frac{t \hat p_0^q}{q_{0,x}} \sum \limits_{j=0}^\alpha \frac{m_{j,x}}{(-z)^{\alpha-j+1}},\\[1mm]
\label{yme-ro1-inv4}
p_\beta^r &= \frac{r_{\beta-1,x}}{q_{0,x}}\hat p_0^q - \hat p_{\beta-1}^r,\\[1mm]
\intertext{resp.}
\label{yme-ro2-inv1}
p_0 &= \frac{\hat p_0^m}{tm_{0,x}},\\[1mm]
\label{yme-ro2-inv2}
p_\alpha^q &= \sum_{j=0}^{\alpha}\frac{(-t)^{{\alpha-j}}\hat p_j^q}{z^{\alpha-j+1}}- \frac{\hat p_{0}^m}{m_{0,x}}\sum_{j=0}^{\alpha}\frac{(-t)^{\alpha-j}q_{j,x}}{z^{\alpha-j+1}},\\[1mm]
\label{yme-ro2-inv3}
p_\alpha^m &=\frac{m_{\alpha+1,x}}{m_{0,x}}\hat p_0^m-\hat p_{\alpha+1}^m,\\[1mm]
\label{yme-ro2-inv4}
p_\beta^r &= -\sum_{j=0}^{\beta-1}\frac{z^{\beta-j-1}\hat p_j^r}{t^{\beta-j}}+\frac{\hat p_0^m}{m_{0,x}}\left(-\frac{z^{\beta}}{t^{\beta+1}}+\sum_{j=1}^{\beta}\frac{z^{\beta-j}r_{j-1,x}}{t^{\beta-j+1}}\right),
\end{align}
where $\alpha\in\mathbb{N}_0, \beta \in\mathbb{N}$, and $\hat p_0^r=-\hat p_0$. 

Moreover, both $({{\mathcal{R}}}^{q})^{-1}$, resp. $({{\mathcal{R}}}^{m})^{-1}$ leave the set $\mathrm{sym}_{\mathcal{A}}^{\tau^W}(\mathcal{E})$ invariant, i.e. they are recursion operators for $\tau^W$-symmetries of the rYME \eqref{yme}.\\
\end{proposition}

\begin{proof}
To prove that ${{\mathcal{R}}}^q$, resp.  ${{\mathcal{R}}}^m$ are bijective, it is enough to verify that $({{\mathcal{R}}}^q)^{-1}$, resp $({{\mathcal{R}}}^m)^{-1}$ are both the left and right inverses of ${{\mathcal{R}}}^q$, resp.  ${{\mathcal{R}}}^m$ on $\mathcal{A}^{\mathbb{N}}$. The proofs of the assertions that $({{\mathcal{R}}}^{q})^{-1}$, resp. $({{\mathcal{R}}}^{m})^{-1}$ leave the set  $\mathrm{sym}_{\mathcal{A}}^{\tau^W}(\mathcal{E})$ invariant can be performed similarly as the proofs of Propositions \ref{yme-ro} and \ref{ROm}.

Since all results can be readily verified by straightforward but tedious computations we omit the details here.
\end{proof}

\begin{corollary}
\label{surj}
The recursion operators $\mathcal{R}^{q}:\mathrm{sym}_{\mathcal{A}}^{\tau^W}(\mathcal{E})\to\mathrm{sym}_{\mathcal{A}}^{\tau^W}(\mathcal{E})$ and $\mathcal{R}^{m}:\mathrm{sym}_{\mathcal{A}}^{\tau^W}(\mathcal{E})\to\mathrm{sym}_{\mathcal{A}}^{\tau^W}(\mathcal{E})$ for $\tau^W$-symmetries of the rYME \eqref{yme} are automorphisms of the real vector space $\mathrm{sym}_{\mathcal{A}}^{\tau^W}(\mathcal{E})$. The inverses $(\mathcal{R}^q)^{-1}$, resp. $(\mathcal{R}^m)^{-1}$ are determined by the formulas \eqref{yme-ro1-inv1} - \eqref{yme-ro1-inv4}, resp. \eqref{yme-ro2-inv1} - \eqref{yme-ro2-inv4}.
\end{corollary}

\begin{remark}
\label{RO-notation}
Of course, the inverse recursion operators $(\mathcal{R}^q)^{-1}$ and $(\mathcal{R}^m)^{-1}$  can also be viewed as infinite-dimensional matrices, see Appendix C. Moreover, it turns out that if we apply $(\mathcal{R}^q)^{-1}$, resp. $(\mathcal{R}^m)^{-1}$ on an arbitrary $\tau^W$-symmetry with a nontrivial local shadow then the shadow of the image, if nonlocal, can depends only on local variables and nonlocal variables from the covering $\tau^q$, reps. from the covering $\tau^m$. Our notation of the recursion operators is based just on this observation.
\end{remark}

From the point of view of the actions of the presented recursion operators on $\tau^W$-symmetries of the rYME \eqref{yme}, it is also useful to know whether they pair-wise commute. The answer is as follows.
\begin{proposition}
\label{commute}
The $\mathcal{A}$-module endomorphisms ${{\mathcal{R}}}^q,{{\mathcal{R}}}^m,({{\mathcal{R}}}^q)^{-1}$ and $({{\mathcal{R}}}^m)^{-1}$ pair-wise commute. 
 \end{proposition}
\begin{proof}
The operators ${{\mathcal{R}}}^q$ and $({{\mathcal{R}}}^q)^{-1}$ commute as they are mutually inverse. The same is true for the operators ${{\mathcal{R}}}^m$ and $({{\mathcal{R}}}^m)^{-1}$. 

To show that the operators ${{\mathcal{R}}}^q$ and $\mathcal{R}^{m}$ commute, we apply their commutator on an arbitrary $P\in{\mathcal{A}}^{\mathbb{N}}$ and use the $\mathcal A$-linearity of $\mathcal R^q$. We obtain
\begin{align*}
[{{\mathcal{R}}}^m,{{\mathcal{R}}}^q](P)&=[z\mathcal{I}+t{{\mathcal{R}}}^q,{{\mathcal{R}}}^q](P)=z\mathcal{I}({{\mathcal{R}}}^q(P))+t{{\mathcal{R}}}^q({{\mathcal{R}}}^q(P))-{{\mathcal{R}}}^q((z\mathcal{I}+t{{\mathcal{R}}}^q)(P))\\
&=z{{\mathcal{R}}}^q(P)+t{{\mathcal{R}}}^q({{\mathcal{R}}}^q(P))-{{\mathcal{R}}}^q(zP)-{{\mathcal{R}}}^q(t{{\mathcal{R}}}^q(P))=0.
\end{align*}

The remaining commutativity relations between $({\mathcal R}^m)^{-1}$ and $\mathcal R^q$, $({\mathcal R}^q)^{-1}$ and $\mathcal R^m$, and $({\mathcal R}^q)^{-1}$ and $({\mathcal R}^m)^{-1}$ successively follow from the equalities
\begin{align*}
\mathcal R^m\circ[({\mathcal R}^m)^{-1}, \mathcal R^q]&=\mathcal R^m \circ \left(({\mathcal R}^m)^{-1} \circ \mathcal R^q-{{\mathcal{R}}}^q\circ({{\mathcal{R}}}^m)^{-1}\right)=\mathcal R^q - \mathcal R^m \circ \mathcal R^q \circ ({\mathcal R}^m)^{-1}\\
&= \mathcal R^q - \mathcal R^q \circ \mathcal R^m \circ ({\mathcal R}^m)^{-1}=\mathcal R^q-\mathcal R^q =0,\\[2mm]
{{\mathcal{R}}}^q\circ[({{\mathcal{R}}}^q)^{-1},{{\mathcal{R}}}^m]&={{\mathcal{R}}}^q\circ\left(({{\mathcal{R}}}^q)^{-1}\circ{{\mathcal{R}}}^m-{{\mathcal{R}}}^m\circ({{\mathcal{R}}}^q)^{-1}\right)={{\mathcal{R}}}^m-{{\mathcal{R}}}^q\circ{{\mathcal{R}}}^m\circ({{\mathcal{R}}}^q)^{-1}\\
&={{\mathcal{R}}}^m-{{\mathcal{R}}}^m\circ{{\mathcal{R}}}^q\circ({{\mathcal{R}}}^q)^{-1}={{\mathcal{R}}}^m-{{\mathcal{R}}}^m =  0,\\[2mm]
{{\mathcal{R}}}^q\circ[({{\mathcal{R}}}^q)^{-1},({{\mathcal{R}}}^m)^{-1}]&={{\mathcal{R}}}^q\circ\left(({{\mathcal{R}}}^q)^{-1}\circ({{\mathcal{R}}}^m)^{-1}-({{\mathcal{R}}}^m)^{-1}\circ({{\mathcal{R}}}^q)^{-1}\right)\\
&=({{\mathcal{R}}}^m)^{-1}-{{\mathcal{R}}}^q\circ({{\mathcal{R}}}^m)^{-1}\circ({{\mathcal{R}}}^q)^{-1}\\
&=({{\mathcal{R}}}^m)^{-1}-({{\mathcal{R}}}^m)^{-1}\circ{{\mathcal{R}}}^q\circ({{\mathcal{R}}}^q)^{-1}=({{\mathcal{R}}}^m)^{-1}-({{\mathcal{R}}}^m)^{-1} = 0,
\end{align*}
and from the fact that $\mathcal R^q$ and $\mathcal R^m$ are injective.
\end{proof}

Hereafter, we will call the recursion operators $\mathcal{R}^q$, $\mathcal{R}^m$ and their inversions $(\mathcal{R}^q)^{-1}$, $(\mathcal{R}^m)^{-1}$ as the \textit{basic recursion operators}, the reason will become clear at the end of this paragraph. The compositions of the basic recursion operators will be denoted as
$$\mathcal R_i^j = \underbrace{({\mathcal R}^q)^{-1} \circ \ldots \circ ({\mathcal R}^q)^{-1}}_{i-times} \circ \underbrace{\mathcal R^m \circ \ldots \circ \mathcal R^m}_{j-times}, \qquad i,j \in \Bbb Z,$$
where for negative values of $i$, resp. $j$, we define
$$\underbrace{({\mathcal R}^q)^{-1} \circ \ldots \circ ({\mathcal R}^q)^{-1}}_{i-times}=\underbrace{\mathcal R^q \circ \ldots \circ \mathcal R^q}_{-i-times}, \quad \mathrm{resp.}\ \underbrace{\mathcal R^m \circ \ldots \circ \mathcal R^m}_{j-times}=\underbrace{({\mathcal R}^m)^{-1} \circ \ldots \circ ({\mathcal R}^m)^{-1}}_{-j-times}.$$
In this notation
$$\mathcal R^q \equiv \mathcal R_{-1}^0, \qquad ({\mathcal R}^q)^{-1} \equiv \mathcal R_1^0, \qquad \mathcal R^m \equiv \mathcal R_0^1, \qquad ({\mathcal R}^m)^{-1} \equiv \mathcal R_0^{-1},\quad \mathcal I \equiv \mathcal R_0^0.$$

It can be easily deduced from Prop. \ref{commute}, that all the the just defined recursion operators $\mathcal{R}_i^j,\ i,j\in\mathbb{Z}$, pair-wise commute, and,  for all $i,j,k,l \in \Bbb Z$ we have
$$\mathcal R_i^j \circ \mathcal R_k^l = \mathcal R_k^l \circ \mathcal R_i^j = \mathcal R_{i+k}^{j+l} \mathrm{\ and\ } (\mathcal R_i^j)^{-1}=\mathcal R_{-i}^{-j}.$$ 
Thus, the set ${\{\mathcal{R}_i^j\}}_{i,j\in\mathbb{Z}}$ carries a commutative group structure with respect to the composition operation, 
the operators $\mathcal{R}_0^1$ and $\mathcal{R}_1^0$ forming the minimal set of its generators.

Moreover, it seems that all the recursion operators for $\tau^W$-symmetries of the rYME arise just as $\mathbb{R}$-linear combinations of the operators from $\{ \mathcal R_i^j \}_{i,j \in \Bbb Z}$. Thus, we \textit{conjecture} that the recursion operators $\mathcal R_{-1}^0, \mathcal R_1^0, \mathcal R_0^1$ and $\mathcal R_0^{-1}$ form the minimal set of generators of the \textit{whole} associative $\mathbb{R}$-algebra of recursion operators for $\tau^W$-symmetries (with respect to the choice $\mathcal{A}=\overline{\mathcal{L}\cup\mathcal{F}}$) of the rYME \eqref{yme}.

 Finally, note that using the defining formula \eqref{yme-ro2} for $\mathcal{R}_0^1\equiv\mathcal{R}^m$, one can simply obtain
 an alternative description of the operators $\mathcal{R}_i^j$:
 $$\mathcal R_i^j = \mathcal R_0^1 \circ \mathcal R_i^{j-1} = (t \mathcal R_{-1}^0 + z \mathcal R_0^0) \circ \mathcal R_i^{j-1} = t \mathcal R_{i-1}^{j-1} + z \mathcal R_i^{j-1}.$$

\section{Actions of the recursion operators on $\tau^W$-symmetries of the rYME}
\label{sec:4}
Once we are given  one or more recursion operators and we know one  `seed' $\tau^W$-symmetry, we are ready to produce a whole hierarchy of $\tau^W$-symmetries for the rYME \eqref{yme}.
The traditional route to achieve this consists in the use of the recursion operators $\mathcal R^q_{sh}$ and $\mathcal R^m_{sh}$ for shadows determined by {relation} \eqref{ro1-shad} and {relation} \eqref{ro2-shad} in the following manner:
\begin{itemize}
\item[1)] We solve the equation \eqref{nsym-yme1} with respect to the unknown (local, if it exists) differential function $p_0$. In other words, we find a (local) shadow $p_0$ of a $\tau^W$-symmetry of the rYME \eqref{yme}.
\item[2)] We apply one of the recursion operators for shadows $\mathcal R^q_{sh}$ and $\mathcal R^m_{sh}$ to the known shadow $p_0$  to produce a new shadow $\hat p_0$. However, applying any of them to a known shadow $p_0$ and trying to find a new shadow $\hat p_0$, we face to a system of two differential equations, and to find its solution can be in general a very non-trivial task, since the solution may depend on nonlocal variables as well. 
\item[3)] We try to lift each shadow $p_0$ at hand to the full-fledged $\tau^W$-symmetry, i.e., having fixed a function $p_0$ that satisfies the condition \eqref{nsym-yme1}, we try to solve Eqs. \eqref{nsym-yme2} - \eqref{nsym-yme9} with respect to unknown functions $p_{\alpha}^q,\ p_{\alpha}^m$ and $p_{\beta}^r$, $\alpha\geq 0,\ \beta\geq 1.$ Note that the lift must be looked for  for each known shadow separately, and moreover, given a shadow, we do not know in advance whether its $\tau^W$-lift exists at all.
\end{itemize} 

However, once we have `full-fledged' recursion operators at our disposal, we are able to obtain the hierarchy of $\tau^W$-symmetries of the rYME \eqref{yme} in a significantly simpler and much more straightforward way: we solve Eqs. \eqref{nsym-yme1} - \eqref{nsym-yme9} only once - in order to obtain a seed  $\tau^W$-symmetry, other new $\tau^W$-symmetries are then generated simply by matrix multiplication.   

Below, the `full-fledged' recursion operators are employed to outline the construction of two (infinite) hierarchies of $\tau^W$-symmetries for the rYME \eqref{yme}. To make the calculations as simple as possible, the seed symmetries $\Psi_0^0$ and $\Omega_0^0$ are chosen to have local shadows there. 
\begin{remark}
 Note that, given a seed symmetry $P$, it is far from obvious that the hierarchy ${\{\mathcal{R}_i^j(P)\}}_{i,j\in\mathbb{Z}}$ is actually infinite. However, the proof of this statement requires a detailed study of the action of the operators on the Lie algebra of symmetries, which is beyond the scope of this paper. We postpone it to a separate article.\end{remark}

\textit{Hierarchy} ${\{\Psi_k^l\}}_{k,l \in \Bbb Z}$.
One can verify by direct computations that the rYME \eqref{yme} admits the $\tau^W$-symmetry
$$\Psi_0^0 = [u_z, \ldots, q_{\alpha,z}, m_{\alpha,z}-(\alpha+1)m_{\alpha+1}, r_{\alpha+1,z}, \ldots], \quad \alpha\geq 0.$$
{Choosing $\Psi_0^0$ as a seed symmetry, we are easily provided with an infinite hierarchy ${\{\Psi_k^l\}}_{k,l \in \Bbb Z}$ of $\tau^W$-symmetries  of the rYME \eqref{yme} with $\Psi_k^l = [\psi_{k,0}^{l},\psi_{k,0}^{l,q},\psi_{k,0}^{l,m},\psi_{k,1}^{l,r},\psi_{k,1}^{l,q},\psi_{k,1}^{l,m},\dots]=\mathcal R_k^l(\Psi_0^0)$.}
In particular, using the matrix representations of the basic recursion operator $\mathcal R_{-1}^0 \equiv \mathcal R^q$ (see Remark \ref{ROq-mat} in Sec. \ref{sec:2}), we can straightforwardly compute the shadow $\psi_{-1,0}^0$ of the $\tau^W$-symmetry $\Psi_{-1}^0$ as follows
$$\psi_{-1,0}^0 = (\mathcal R_{-1}^0 \cdot (\Psi_0^0)^T)_1 = u_x u_z+r_{1,z} = u_x u_z - u_x u_z + u_t = u_t.$$
Further, one obtains {the} components $\psi_{-1,\alpha}^{0,q}, \psi_{-1,\alpha}^{0,m}, \psi_{-1,\alpha+1}^{0,r}$, $\alpha \geq 0$, of $\Psi_{-1}^0$ in the form:
\begin{align*}
\psi_{-1,\alpha}^{0,q} &= (\mathcal R_{-1}^0 \cdot (\Psi_0^0)^T)_{3\alpha+2}=q_{\alpha,x}u_z+q_{\alpha-1,z} = q_{\alpha,t},\\[2mm]
\psi_{-1,\alpha}^{0,m} &= (\mathcal R_{-1}^0 \cdot (\Psi_0^0)^T)_{3\alpha+3}=m_{\alpha,x}u_z-\frac{1}{t} (m_{\alpha-1,z}-\alpha m_{\alpha})-\frac{z}{t}(m_{\alpha,z}-(\alpha+1)m_{\alpha+1})\\
& = m_{\alpha,x}u_z - \frac{1}{t} m_{\alpha-1,z} - \frac{z}{t} m_{\alpha,z} + \frac{\alpha}{t}m_{\alpha} + \frac{(\alpha+1)z}{t}m_{\alpha+1}\\
& = m_{\alpha,t} + \frac{\alpha}{t}m_{\alpha} + \frac{(\alpha+1)z}{t}m_{\alpha+1}\\[2mm]
\psi_{-1,\alpha+1}^{0,r} & = (\mathcal R_{-1}^0 \cdot (\Psi_0^0)^T)_{3\alpha+4} = r_{\alpha+1,x}u_z-r_{\alpha+2,z} =  r_{\alpha+1,x}u_z-r_{\alpha+1,x}u_z + r_{\alpha+1,t} = r_{\alpha+1,t}.
\end{align*}
{Thus, the new $\tau^W$-symmetry $\Psi_{-1}^0$ takes the form}
\begin{align*}
\Psi_{-1}^0&= \mathcal R_{-1}^0(\Psi_0^0)=\left[u_t, \ldots, q_{\alpha,t}, m_{\alpha,t}+\frac{\alpha}{t} m_\alpha+\frac{(\alpha+1)z}{t}m_{\alpha+1}, r_{\alpha+1,t}, \ldots\right], \; \alpha\geq 0.
\end{align*}

{Let us point out that there is another $\tau^W$-symmetry with a local shadow within the hierarchy${\{\Psi_k^l\}}_{k,l \in \Bbb Z}$, namely}
\begin{align*}
\Psi_0^1&= R_0^1(\Psi_0^0)=t\Psi_{-1}^0+z\Psi_0^0,
\end{align*}
the shadow being $\psi_{0,0}^1= t\psi_{-1,0}^0+z\psi_{0,0}^0 = tu_t+zu_z$.

{In Tab. \ref{tab1}, we present nonlocal shadows of other selected $\tau^W$-symmetries which belong to the hierarchy ${\{\Psi_k^l\}}_{k,l \in \Bbb Z}$}.\\
\begin{table}[h]
$\begin{array}{c | c}
\tau^W{\text -symmetry}\ \Psi_k^l& Shadow\ \psi_{k,0}^l\\[1mm]
\hline\\[-2mm]
\Psi_{-2}^0 = \mathcal R_{-2}^0(\Psi_0^0) & \psi_{-2,0}^0=r_{1,t}+u_tu_x\\[2mm]
\Psi_{-3}^0 = \mathcal R_{-3}^0(\Psi_0^0) & \psi_{-3,0}^0=-r_{2,t}+u_x r_{1,t} + u_tu_y\\[2mm]
\Psi_1^0 = \mathcal R_{1}^0(\Psi_0^0) & \displaystyle \psi_{1,0}^0=\frac{q_{0,z}}{q_{0,x}}\\[5mm]
\Psi_{2}^0 = \mathcal R_{2}^0(\Psi_0^0) & \displaystyle \psi_{2,0}^0=\frac{q_{1,z}}{q_{0,x}}-\frac{q_{0,z} q_{1,x}}{q_{0,x}^2}\\[4mm]
\Psi_0^2 = \mathcal R_0^2(\Psi_0^0) & \psi_{0,0}^2=t^2(r_{1,t}+u_tu_x)+2tzu_t+z^2u_z\\[2mm]
\Psi_0^3 = \mathcal R_0^3(\Psi_0^0) & \psi_{0,0}^3=t^3(-r_{2,t}+u_x r_{1,t} + u_tu_y)+3t^2z(r_{1,t}+u_tu_x)+3tz^2u_t+z^3u_z\\[2mm]
\Psi_0^{-1} = \mathcal R_0^{-1}(\Psi_0^0) & \psi_{0,0}^{-1}=\displaystyle \frac{m_{0,z}-m_1}{tm_{0,x}}\\[5mm]
\Psi_0^{-2} = \mathcal R_0^{-2}(\Psi_0^0) & \psi_{0,0}^{-2}=\displaystyle \frac{2m_2-m_{1,z}}{t m_{0,x}}+\frac{m_{1,x}(m_{0,z}-m_1)}{t m_{0,x}^2}
\end{array}$\\[2mm]
\caption{Shadows of selected $\tau^W$-symmetries from the hierarchy ${\{\Psi_k^l\}}_{k,l \in \Bbb Z}$ \label{tab1}}
\end{table}

\textit{Hierarchy} ${\{\Omega_k^l\}}_{k,l \in \Bbb Z}$.
Another $\tau^W$-symmetry of the rYME \eqref{yme} with a local shadow is $$\Omega_0^0 =\left[\omega_{0,0}^0,\omega_{0,0}^{0,q},\omega_{0,0}^{0,m},\omega_{0,1}^{0,r}, \omega_{0,1}^{0,q},\omega_{0,1}^{0,m},\omega_{0,2}^{0,r}, \ldots  \right],$$ with
\begin{align*}
\omega_{0,0}^0&=tu_t-xu_x+2u, \qquad \omega_{0,\alpha}^{0,q} = tq_{\alpha,t}-xq_{\alpha,x}-\alpha q_\alpha,\\[2mm]
\omega_{0,\alpha}^{0,m} & = tm_{\alpha,t}-xm_{\alpha,x}, \qquad \omega_{0,\beta}^{0,r}=tr_{\beta,t}-xr_{\beta,x}+(\beta+2)r_\beta,\quad \alpha\geq 0, \beta \geq 1.
\end{align*}
Choosing it as a seed symmetry, we arrive at the hierarchy  ${\{\Omega_k^l\}}_{k,l \in \Bbb Z}$ of $\tau^W$-symmetries, where $$\Omega_k^l=[\omega_{k,0}^{l},\omega_{k,0}^{l,q},\omega_{k,0}^{l,m},\omega_{k,1}^{l,r},\omega_{k,1}^{l,q},\omega_{k,1}^{l,m},\dots]=\mathcal{R}_k^l(\Omega_0^0).$$ 
Within this hierarchy, there are also two other symmetries with local shadows, namely
\begin{align*}
\Omega_1^0&= \mathcal R_1^0(\Omega_0^0)=[tu_z-x, \ldots, tq_{\alpha,z}-(\alpha+1)q_{\alpha+1}, tm_{\alpha,z}, tr_{\alpha+1,z}-(\alpha+2)r_{\alpha}, \ldots], \; \alpha\geq 0,
\intertext{and}
\Omega_1^1&=\mathcal R_0^1(\Omega_1^0)=t\, \Omega_0^0+z\,\Omega_1^0
\end{align*}
with the shadow $\omega_{1,0}^1= t\omega_{0,0}^0+z{\omega_{1,0}^0} = t(tu_t-xu_x+2u)+z(tu_z-x)$.

In Tab. \ref{tab2}, we present nonlocal shadows of selected $\tau^W$-symmetries from the hierarchy ${\{\Omega_k^l\}}_{k,l \in \Bbb Z}$.
\begin{table}[h]
$\begin{array}{c | c}
\tau^W{\text -symmetry}\ \Omega_k^l& Shadow\ \omega_{k,0}^l\\[1mm]
\hline\\[-2mm]
\Omega_{-1}^0 = \mathcal R_{-1}^0(\Omega_0^0) & \omega_{-1,0}^0=3r_1+2uu_x+t(u_tu_x+r_{1,t})-xu_y\\[2mm]
\Omega_{-2}^0 = \mathcal R_{-2}^0(\Omega_0^0) & \omega_{-2,0}^0=-4r_2+3u_xr_1+2uu_y+t(u_tu_y+u_xr_{1,t}-r_{2,t})-x(u_xu_y+r_{1,y})\\[2mm]
\Omega_2^0 = \mathcal R_{2}^0(\Omega_0^0) & \displaystyle \omega_{2,0}^0=\frac{tq_{0,z}-q_1}{q_{0,x}}\\[5mm]
\Omega_{3}^0 = \mathcal R_{3}^0(\Omega_0^0) & \displaystyle \omega_{3,0}^0=\frac{tq_{1,z}-2q_2}{q_{0,x}}-\frac{q_{1,x}(tq_{0,z}-q_1)}{q_{0,x}^2}\\[4mm]
\Omega_0^1 = \mathcal R_0^1(\Omega_0^0) & \omega_{0,0}^1=t(3r_1+2uu_x+tu_tu_x+tr_{1,t}-xu_y)+z(tu_t-xu_x+2u)\\[2mm]
\Omega_0^{-1} = \mathcal R_0^{-1}(\Omega_0^0) & \displaystyle \omega_{0,0}^{-1}=u_z-\frac{x}{t}-\frac{zm_{0,z}}{tm_{0,x}}\\[5mm]
\Omega_0^{-2} = \mathcal R_0^{-2}(\Omega_0^0) & \displaystyle \omega_{0,0}^{-2}=\frac{zm_{1,z}+m_{0,z}}{tm_{0,x}}-\frac{zm_{1,x}m_{0.z}}{tm_{0,x}^2}\\[5mm]
\Omega_{1}^{-1} = \mathcal R_{1}^{-1}(\Omega_0^{0}) & \displaystyle \omega_{1,0}^{-1}=\frac{m_{0,z}}{m_{0,x}}
\end{array}$\\[2mm]
\caption{Shadows of selected $\tau^W$-symmetries from the hierarchy $\{\Omega_k^l\}_{k,l \in \Bbb Z}$ \label{tab2}}
\end{table}

The rYME \eqref{yme} admits also families of $\tau^W$-symmetries indexed by arbitrary functions $A=A(z,q_0)$, $B=B(t,y)$, resp. $C=C(z/t,m_0)$ (for brevity, we refer to these families as to $\mathrm{Fam}_A,\ \mathrm{Fam}_B,$ and $\mathrm{Fam}_C$, respectively), however, the complete description of these families is a rather complex matter. In the three examples below, we give particular instances of such $\tau^W$-symmetries and show how the action of the recursion operators $\mathcal{R}_i^j$ on them may help us to describe at least some of the symmetries within the corresponding family. 
The common feature of all the examples is the following:
we apply a `full-fledged' recursion operator to a particularly chosen full-fledged $\tau^W$-symmetry from $\mathrm{Fam}_A$, $\mathrm{Fam}_B$, resp. $\mathrm{Fam}_C$, which \textit{immediately} provides us with another $\tau^W$-symmetry that lies in the same family as the original one. On the contrary, if we knew only the shadow (still depending on an arbitrary function $A$, $B$, resp. $C$) of the latter and tried to lift it and discover a concise formula describing all its components directly, we would have to go through lots of tedious calculations, the success being unwarranted. Thus, the benefits of having a `full-fledged' recursion operator at our disposal become evident.

\begin{example}
\label{ex:1}
{\color{red}}The direct computations show that one of the $\tau^W$-symmeties of the rYME \eqref{yme} belonging to $\mathrm{Fam}_B$ mentioned above is
$$\Upsilon_0^0(B) = \left[\upsilon_{0,0}^0(B), 0,0,\upsilon_{0,1}^{0,r}(B), 0,0, \upsilon_{0,2}^{0,r}(B),0,0, \upsilon_{0,3}^{0,r}(B), \ldots  \right],$$
where $\upsilon_{0,0}^0(B)=B$, and its other nontrivial components $\upsilon_{0,\beta}^{0,r}(B)$, $\beta\geq 1$ can be quite easily  described in a concise form by means of the operator 
$$\mathfrak V = z\pd{}{t}+x\pd{}{y}-2u\pd{}{x}+3r_1\pd{}{u}-\sum\limits_{i=1}^\infty (i+3)r_{i+1}\pd{}{r_i},$$
which yields
the relations
$$\upsilon_{0,\beta}^{0,r}(B) =\frac{(-1)^{\beta-1}}{\beta!}\mathfrak V^\beta(\upsilon_{0,0}^{0}(B))=\frac{(-1)^{\beta-1}}{\beta!}\mathfrak V^\beta(B), \quad \beta \geq 1,$$
where $\mathfrak V^\beta = \underbrace{\mathfrak V \circ \ldots \circ \mathfrak V}_{\beta-times}$.  

Applying the basic recursion operator $\mathcal R_{-1}^0$ on $\Upsilon_0^0(B)$, one can directly obtain the $\tau^W$-symmetry
$$\Upsilon_{-1}^0(B) = \left[\upsilon_{-1,0}^0(B), \upsilon_{-1,0}^{0,q}(B),\upsilon_{-1,0}^{0,m}(B),\upsilon_{-1,1}^{0,r}(B), \upsilon_{-1,1}^{0,q}(B),\upsilon_{-1,1}^{0,m}(B), \upsilon_{-1,2}^{0,r}(B), \ldots  \right],$$
where for each $\alpha \geq 0$, $\beta \geq 1$ we have
\begin{align*}
\upsilon_{-1,0}^0(B) &= u_x \upsilon_{0,0}^{0}(B) + \upsilon_{0,1}^{0,r}(B) = u_xB + zB_t + xB_y,\\[2mm]
\upsilon_{-1,\alpha}^{0,q}(B) &= q_{\alpha,x} \upsilon_{0,0}^{0}(B) = q_{\alpha,x} B,\\[2mm]
\upsilon_{-1,\alpha}^{0,m}(B) &= m_{\alpha,x} \upsilon_{0,0}^{0}(B) = m_{\alpha,x} B,\\[2mm]
\intertext{and}
\upsilon_{-1,\beta}^{0,r}(B) &= r_{\beta,x} \upsilon_{0,0}^{0}(B)  - \upsilon_{0,\beta+1}^{0,r}(B) = r_{\beta,x} B - \frac{(-1)^\beta}{(\beta+1)!} \mathfrak V^{\beta+1}(B).
\end{align*}

Let us remark that both the first column of the matrix $\mathcal R^q$ (see Sec. \ref{sec:2}) and the first column of the matrix $\mathcal R^m$ (see Appendix C) coincide with the symmetry $\Upsilon_{-1}^0(B)$,  the first one upon putting $B(t,y) \equiv 1$, the second one upon putting $B(t,y)=t$. 
\end{example}

\begin{example}
\label{ex:2}
One of the $\tau^W$-symmetries of the rYME \eqref{yme} lying in the family $\mathrm{Fam}_A$ is the following invisible $\tau^W$-symmetry:
 $$\Xi_0^0(A) = \left[0, \xi_{0,0}^{0,q}(A),0,0, \xi_{0,1}^{0,q}(A),0,0,\xi_{0,2}^{0,q}(A),0,0, \ldots  \right],$$
where $\xi_{0,0}^{0,q}(A)=A$, and the components $ \xi_{0,\alpha}^{0,q}(A)$, $\alpha \geq 1$, are described by the formula 
$$\xi_{0,\alpha}^{0,q}(A) = \frac{1}{\alpha}\mathfrak X(\xi_{0,\alpha-1}^{0,q}(A)) = \frac{1}{\alpha !}\mathfrak X^\alpha(\xi_{0,0}^{0,q}(A)) = \frac{1}{\alpha !}\mathfrak X^\alpha(A),$$
where
$$\mathfrak X = t\pd{}{z}+\sum \limits_{i=0}^{\infty} (i+1)q_{i+1}\pd{}{q_i}.$$

The action of the basic recursion operator $\mathcal R_1^0$ on  $\Xi_0^0(A)$ leads to the symmetry 
$$\Xi_1^0(A) = \left[\xi_{1,0}^0(A), \xi_{1,0}^{0,q}(A),\xi_{1,0}^{0,m}(A),\xi_{1,1}^{0,r}(A), \xi_{1,1}^{0,q}(A),\xi_{1,1}^{0,m}(A), \xi_{1,2}^{0,r}(A), \ldots  \right],$$ 
where for each $\alpha \geq 0$, $\beta \geq 1$ we have
\begin{align*}
\xi_{1,0}^{0}(A) &= \frac{\xi_{0,0}^{0,q}(A)}{q_{0,x}}=\frac{A}{q_{0,x}},\\
\xi_{1,\alpha}^{0,q}(A) &= -\frac{q_{\alpha+1,x}}{q_{0,x}}\cdot \xi_{0,0}^{0,q}(A) + \xi_{0,\alpha+1}^{0,q}(A)= -\frac{q_{\alpha+1,x}}{q_{0,x}}\cdot A  + \frac{1}{(\alpha+1)!}\mathfrak{X}^{\alpha+1}(A),\\
\xi_{1,\alpha}^{0,m}(A) &=-\frac{t \cdot \xi_{0,0}^{0,q}(A)}{q_{0,x}} \cdot \sum \limits_{j=0}^\alpha \frac{m_{j,x}}{(-z)^{\alpha-j+1}}=-\frac{t A}{q_{0,x}} \cdot \sum \limits_{j=0}^\alpha \frac{m_{j,x}}{(-z)^{\alpha-j+1}}\\
\xi_{1,\beta}^{0,q}(A)&=\frac{r_{\beta-1,x}}{q_{0,x}}\cdot \xi_{0,0}^{0,q}(A) =\frac{r_{\beta-1,x}}{q_{0,x}}\cdot A.
\end{align*}

Note that upon putting $A(z,q_0) \equiv 1$, we obtain the $\tau^W$-symmetry $\Xi_{1}^0(1)$ which coincides with the second column of the matrix $(\mathcal R^q)^{-1}$ (see Appendix C).
\end{example}

\begin{example} 
\label{ex:3}
An example of a $\tau^W$-symmetry of the rYME \eqref{yme} depending on an arbitrary function $C=C(z/t,m_0)$ is an invisible $\tau^W$-symmetry
$$\Theta_0^0(C) = \left[0,0,\theta_{0,0}^{0,m}(C), 0,0, \theta_{0,0}^{0,m}(C),0,0, \theta_{0,0}^{0,m}(C), 0,0,\ldots  \right],$$
where $\theta_{0,0}^{0,m}(C)=C$, and the components $\theta_{0,\alpha}^{0,m}(C)$, $\alpha \geq 1$ are 
given by the formula
$$\theta_{0,\alpha}^{0,m}(C) = \frac{1}{\alpha}\mathfrak T(\theta_{0,\alpha-1}^{0,m}(C)) = \frac{1}{\alpha !}\mathfrak T^\alpha(\theta_{0,0}^{0,m}(C)) = \frac{1}{\alpha !}\mathfrak T^\alpha(C),$$
where
$$\mathfrak T = \pd{}{z}+\sum\limits_{i=0}^\infty (i+1) m_{i+1}\pd{}{m_i}.$$

Applying the basic recursion operator $\mathcal R_0^{-1}$ on $\Theta_0^0(C)$, we obtain the $\tau^W$-symmetry
$$\Theta_0^{-1}(C) = \left[\theta_{0,0}^{-1}(C), \theta_{0,0}^{-1,q}(C),\theta_{0,0}^{-1,m}(C),\theta_{0,1}^{-1,r}(C), \theta_{0,1}^{-1,q}(C),\theta_{0,1}^{-1,m}(C), \theta_{0,2}^{-1,r}(C), \ldots  \right],$$
its components for all $\alpha \geq 0$, $\beta \geq 1$ being given by the formulas
\begin{align*}
\theta_{0,0}^{-1}(C) &= \frac{\theta_{0,0}^{0,m}(C)}{tm_{0,x}} = \frac{C}{tm_{0,x}},\\[2mm]
\theta_{0,\alpha}^{-1,q}(C) &= - \frac{\theta_{0,0}^{0,m}(C)}{m_{0,x}}\sum_{j=0}^{\alpha}\frac{(-t)^{\alpha-j}q_{j,x}}{z^{\alpha-j+1}} = - \frac{C}{m_{0,x}}\sum_{j=0}^{\alpha}\frac{(-t)^{\alpha-j}q_{j,x}}{z^{\alpha-j+1}},\\[2mm]
\theta_{0,\alpha}^{-1,m}(C) &= \frac{m_{\alpha+1,x}}{m_{0,x}} \theta_{0,0}^{0,m}(C) -\theta_{0,\alpha+1}^{0,m}(C) = \frac{m_{\alpha+1,x}}{m_{0,x}} \cdot C -\frac{1}{(\alpha+1)!} \mathfrak T^{\alpha+1}(C),\\[2mm]
\intertext{and}
\theta_{0,\beta}^{-1,r}(C) &= \frac{\theta_{0,0}^{0,m}(C)}{m_{0,x}}\left(-\frac{z^{\beta}}{t^{\beta+1}}+\sum_{j=1}^{\beta}\frac{z^{\beta-j}r_{j-1,x}}{t^{\beta-j+1}}\right) = \frac{C}{m_{0,x}}\left(-\frac{z^{\beta}}{t^{\beta+1}}+\sum_{j=1}^{\beta}\frac{z^{\beta-j}r_{j-1,x}}{t^{\beta-j+1}}\right).
\end{align*}

It appears that the $\tau^W$-symmetry $\Theta_0^{-1}(1)$ that lies in the just discussed hierarchy corresponding to the function $C(t/z,m_0)\equiv 1$ coincides with the third column of the matrix $(\mathcal R^m)^{-1}$ (see Appendix C).
\end{example}

\section{Conclusions and future prospects}
In the current literature, recursion operators for nonlocal symmetries of integrable systems are traditionally presented as {\textit{relations} between their shadows only}. {In principle, knowing} such a recursion operator and one shadow, we are able to generate (infinite) hierarchy of new shadows. However, given a differential covering $\tau$ of the equation in question, in case we are interested not only in the shadows but in the full-fledged $\tau$-nonlocal symmetries, we have to investigate separately for each shadow whether it can be lifted to a full-fledged $\tau$-nonlocal symmetry or not. Moreover, 
another disadvantage of the recursion operators for shadows is that one shadow can be related to more than one (possibly to infinitely many) shadows, since the trivial shadow can be (and usually is) related to a nontrivial one.
From this point of view,  recursion operators for shadows can be understood to provide us with a `highly multivalued mapping' between full-fledged symmetries. Thus, the difficulties in using such operators to construct an infinite hierarchy of full-fledged $\tau$-nonlocal symmetries become evident.

In this paper, we have shown that the expansion of the Lax pair of a given Lax-integrable equation may provide us with a differential covering which subsequently enables us to find a \textit{mapping} that puts full-fledged (in this covering) nonlocal symmetries to other ones. This full-fledged recursion operator allows us to generate a (possibly infinite) hierarchy of full-fledged nonlocal symmetries from one known seed full-fledged symmetry directly, without any problem with lifting of shadows. Even though the idea of how to find a full-fledged recursion operator is quite straightforward, the implementation of the latter is not simple at all. The main difficulty of this construction consists in the fact that the covering we have used is infinite-dimensional, and the generating functions of symmetries are in fact infinite sequences of nonlocal functions. Since the recursion operator we are looking for is to act on an infinite sequence resulting in another infinite sequence, our success depends on whether we are able to find general formulas describing the relationships between all the components of the pre-image and its image. We do not claim that this problem can always be satisfactorily overcome, however, we have demonstrated on a particular example of the reduced quasi-classical self-dual Yang-Mills equation that at least sometimes it is definitely possible.

As we have shown in Secs. \ref{sec:2} and \ref{sec:3}, the rYME \eqref{yme} admits two basic pairs of mutually inverse full-fledged recursion operators, such that all of them pair-wise commute. Moreover, it is possible to represent all these recursion operators by infinite-dimensional matrices of differential functions where each matrix has only finitely many nonzero entries in each of its rows, the actions of such recursion operators on full-fledged nonlocal symmetries being then simply given as matrix multiplications without any convergence troubles. This enables us to easily generate infinite hierarchies of full-fledged nonlocal symmetries from known ones, see Sec. \ref{sec:4}. The use of  a `full-fledged' recursion operator turns out to be particularly advantageous when we let it act on full-fledged symmetries depending on arbitrary functions and generate hierarchies thereof, see Examples \ref{ex:1} - \ref{ex:3}. Nevertheless, a more detailed description of such hierarchies is beyond the scope of this paper and we postpone it, as well as the investigation of the Lie algebra structure on the set of  $\tau^W$-symmetries of the rYME \eqref{yme} and a more detailed study of the action of the full-fledged recursion operators on it for a forthcoming separate paper. 

Based on our preliminary computations, we are pretty sure that the full-fledged recursion operators can be similarly constructed and described also for other linearly degenerate (in the sense of \cite{FerMoss}) Lax integrable equations. As already mentioned in the introductory section of this paper, no serious problems in this direction are expected for equations from Tab. 2 in \cite{DFKN}. 

Note that recursion operators for full-fledged $\tau$-nonlocal symmetries of the rYME \eqref{yme} similar to $\mathcal{R}^q$ and $\mathcal{R}^m$ could be in fact constructed also in the `poorer' covering $\tau=\tau^q\oplus\tau^r$. However, the lack of nonlocal variables would make the mapping $\mathcal{R}^m$ not surjective, thus, the hierarchies of nonlocal symmetries obtained from one seed symmetry would not be so large. This situation occurs for example in the case we consider the $4$D universal hierarchy equation
\begin{equation*}u_{yz} = u_{tt}-u_z u_{tx}+u_t u_{xz}\end{equation*}
from Tab. 3 in \cite{DFKN}, cf. also \cite{bog, kramor}, 
since it is unclear to us at the moment, given its multi-parameter Lax pair, how one could derive a covering which would be analogous to the $\tau^m$-covering \eqref{yme-taum}.  

Finally, let us remark that it would be interesting to find out whether the construction of a full-fledged recursion operator similar to the presented one can be performed also for an equation that is not linearly degenerate, for example, for one of the equations that belong to the class of Monge-Amp\`ere type equations, see e.g. \cite{DFKN35, DFKN} and references therein for more details and examples of such equations.

\section*{Acknowledgments}
\label{sec:acknowledgments}

The symbolic computations were performed using the software \textsc{Jets} \cite{Jets}. Computational resources were provided by the e-INFRA CZ project (ID:90254), supported by the Ministry of Education, Youth and Sports of the Czech Republic. The research was supported by the Ministry of Education, Youth and Sports of the Czech Republic (MSMT CR) under RVO funding for IC47813059.\\

\section*{Appendix A: Proof of Proposition \ref{yme-ro}}
We need to check that all components $\hat p_0, \hat p_{\alpha}^q, \hat p_{\alpha}^m, \hat p_{\beta}^r,$ $\alpha\in \Bbb N_0,$ $\beta \in~\Bbb N$, of $\hat P$ satisfy the corresponding conditions \eqref{nsym-yme1}-\eqref{nsym-yme9} on the equation $\tilde{\mathcal{E}}^W$, provided that $p_0,p_{\alpha}^q,p_{\alpha}^m,p_{\beta}^r$ satisfy them.
\begin{itemize}
\item[1)] One obtains by straightforward computation:
\begin{align*}
\tilde D_{yz}(\hat p_0) &-\tilde D_{tx}(\hat p_0) +u_z \tilde D_{xx}(\hat p_0)-u_x\tilde D_{xz}(\hat p_0)+u_{xx}\tilde D_z(\hat p_0)-u_{xz}\tilde D_x(\hat p_0)\\
&=p_0\tilde{D}_x(\underbrace{u_{yz}-u_{tx}+u_zu_{xx}-u_xu_{xz}}_{=\, 0 \ \mathrm{due\ to\ Eq.}\eqref{yme}})\\
&+(\tilde D_y-u_x\tilde D_x+u_{xx})(\underbrace{\tilde D_z(p_1^r)-\tilde D_t(p_0)+u_x\tilde D_z(p_0)+u_z\tilde D_x(p_0)}_{=\, 0 \ \mathrm{due\ to\ Eq.}\eqref{nsym-yme7}})\\
&-(\tilde D_t-u_z\tilde D_x+u_{xz})(\underbrace{\tilde{D}_x(p_1^r)-\tilde D_y(p_0)+2u_x\tilde D_x(p_0)}_{=\, 0 \ \mathrm{due\ to\ Eq.}\eqref{nsym-yme6}}) =0,
\end{align*}
thus $\hat p_0$ satisfies the condition \eqref{nsym-yme1} on the equation $\tilde{\mathcal{E}}^W$. This means that $\hat p_0$ is a shadow of nonlocal symmetry od the rYME \eqref{yme}.\\

\item[2)] For each $\alpha \in \Bbb N_0$, we have (recall that $q_j = p_j^q=0$ if $j<0$):
\begin{align*}
\tilde D_t(\hat p_{\alpha}^q)&-u_z\tilde D_x(\hat p_{\alpha}^q)-q_{\alpha,x}\tilde D_z(\hat p_0)-\tilde D_z(\hat p_{\alpha-1}^q)\\
&=p_0\tilde D_x(\underbrace{q_{\alpha,t}-u_zq_{\alpha,x}-q_{\alpha-1,z}}_{=\, 0 \ \mathrm{due\ to\ Eq.}\eqref{yme-tauq}})\\
&-q_{\alpha,x}(\underbrace{\tilde D_z(p_1^r)-\tilde D_t(p_0)+u_x\tilde D_z(p_0)+u_z\tilde D_x(p_0)}_{=\, 0 \ \mathrm{due\ to\ Eq.}\eqref{nsym-yme7}})\\
&+(\underbrace{\tilde D_t(p_{\alpha-1}^q)-u_z\tilde D_x(p_{\alpha-1}^q)-q_{\alpha-1,x}\tilde D_z(p_0)-\tilde D_z(p_{\alpha-2}^q)}_{=\, 0 \ \mathrm{due\ to\ Eq.}\eqref{nsym-yme2}})=0,
\intertext{and also}
\tilde D_y(\hat p_{\alpha}^q)&-u_x\tilde D_x(\hat p_{\alpha}^q)-q_{\alpha,x}\tilde D_x(\hat p_0)-\tilde D_x(\hat p_{\alpha-1}^q)\\
&=p_0\tilde D_x(\underbrace{q_{\alpha,y}-u_xq_{\alpha,x}-q_{\alpha-1,x}}_{=\, 0\ \mathrm{due\ to\ Eq.}\eqref{yme-tauq}})\\
&-q_{\alpha,x}(\underbrace{\tilde D_x(p_1^r)-\tilde D_y(p_0)+2u_x\tilde D_x(p_0)}_{=\, 0 \ \mathrm{due\ to\ Eq.}\eqref{nsym-yme6}})\\
&+(\underbrace{\tilde D_y(p_{\alpha-1}^q)-u_x\tilde D_x(p_{\alpha-1}^q)-q_{\alpha-1,x}\tilde D_x(p_0)-\tilde D_x(p_{\alpha-2}^q)}_{=\, 0 \ \mathrm{due\ to\ Eq.}\eqref{nsym-yme3}})=0,
\end{align*}
thus $\hat p_\alpha^q$ satisfies the conditions \eqref{nsym-yme2} and \eqref{nsym-yme3} on the equation $\tilde{\mathcal{E}}^W$.\\

\item[3)] For each $\alpha \in \Bbb N_0$, we have (recall that $m_j = p_j^m=0$ if $j<0$):
\begin{align*}
\tilde D_t(\hat p_{\alpha}^m)&-u_z\tilde D_x(\hat p_{\alpha}^m)-m_{\alpha,x}\tilde D_z(\hat p_0)+\frac zt\tilde D_z(\hat p_{\alpha}^m)+\frac 1t\tilde D_z(\hat p_{\alpha-1}^m)\\
&=p_0\tilde D_x\underbrace{\left(m_{\alpha,t}-u_zm_{\alpha,x}+\frac ztm_{\alpha,z}+\frac 1tm_{\alpha-1,z}\right)}_{=\, 0 \ \mathrm{due\ to\ Eq.}\eqref{yme-taum}}\\
&-m_{\alpha,x} (\underbrace{\tilde D_z(p_1^r)-\tilde D_t(p_0)+u_x\tilde D_z(p_0)+u_z\tilde D_x(p_0)}_{=\, 0 \ \mathrm{due\ to\ Eq.}\eqref{nsym-yme7}})\\
&-\frac 1t \underbrace{\left(\tilde D_t(p_{\alpha-1}^m)-u_z\tilde D_x(p_{\alpha-1}^m)-m_{\alpha-1,x}\tilde D_z(p_0)+\frac zt\tilde D_z(p_{\alpha-1}^m)+\frac 1t\tilde D_z(p_{\alpha-2}^m)\right)}_{=\, 0 \ \mathrm{due\ to\ Eq.}\eqref{nsym-yme4}}\\
&-\frac zt \underbrace{\left(\tilde D_t(p_{\alpha}^m)-u_z\tilde D_x(p_{\alpha}^m)-m_{\alpha,x}\tilde D_z(p_0)+\frac zt\tilde D_z(p_{\alpha}^m)+\frac 1t\tilde D_z(p_{\alpha-1}^m)\right)}_{=\, 0 \ \mathrm{due\ to\ Eq.}\eqref{nsym-yme4}} = 0,
\intertext{and also}
\tilde D_y(\hat p_{\alpha}^m)&-u_x\tilde D_x(\hat p_{\alpha}^m)-m_{\alpha,x}\tilde D_x(\hat p_0)+\frac zt\tilde D_x(\hat p_{\alpha}^m)+\frac 1t\tilde D_x(\hat p_{\alpha-1}^m)\\
&=p_0\tilde D_x\underbrace{\left(m_{\alpha,y}-u_xm_{\alpha,x}+\frac ztm_{\alpha,x}+\frac 1tm_{\alpha-1,x}\right)}_{=\, 0 \ \mathrm{due\ to\ Eq.}\eqref{yme-taum}}\\
&-m_{\alpha,x} (\underbrace{\tilde D_x(p_1^r)-\tilde D_y(p_0)+2u_x\tilde D_x(p_0)}_{=\, 0 \ \mathrm{due\ to\ Eq.}\eqref{nsym-yme6}})\\
&-\frac 1t \underbrace{\left(\tilde D_y(p_{\alpha-1}^m)-u_x\tilde D_x(p_{\alpha-1}^m)-m_{\alpha-1,x}\tilde D_x(p_0)+\frac zt\tilde D_x(p_{\alpha-1}^m)+\frac 1t\tilde D_x(p_{\alpha-2}^m)\right)}_{=\, 0 \ \mathrm{due\ to\ Eq.}\eqref{nsym-yme5}}\\
&-\frac zt \underbrace{\left(\tilde D_y(p_{\alpha}^m)-u_x\tilde D_x(p_{\alpha}^m)-m_{\alpha,x}\tilde D_x(p_0)+\frac zt\tilde D_x(p_{\alpha}^m)+\frac 1t\tilde D_x(p_{\alpha-1}^m)\right)}_{=\, 0 \ \mathrm{due\ to\ Eq.}\eqref{nsym-yme5}} = 0,
\end{align*}
thus $\hat p_\alpha^m$ satisfies the conditions \eqref{nsym-yme4} and \eqref{nsym-yme5} on the equation $\tilde{\mathcal{E}}^W$.\\

\item[4)] For the quantity $\hat p_1^r$ we obtain:
\begin{align*}
\tilde D_x(\hat p_1^r)&-\tilde D_y(\hat p_0)+2u_x\tilde D_x(\hat p_0)\\
&=p_0\tilde D_x(\underbrace{r_{1,x}+u_x^2-u_y}_{=\, 0 \ \mathrm{due\ to\ Eq.}\eqref{yme-taur}})+u_x (\underbrace{\tilde D_x(p_1^r)-\tilde D_y(p_0)+2u_x\tilde D_x(p_0)}_{=\, 0 \ \mathrm{due\ to\ Eq.}\eqref{nsym-yme6}})\\
&-(\underbrace{\tilde D_x(p_2^r)-u_x\tilde D_x(p_1^r)-r_{1,x}\tilde D_x(p_0)+\tilde D_y(p_1^r)}_{=\, 0 \ \mathrm{due\ to\ Eq.}\eqref{nsym-yme8}}) =0,
\intertext{and also}
\tilde D_z(\hat p_1^r)&-\tilde D_t(\hat p_0)+u_x\tilde D_z(\hat p_0)+u_z\tilde D_x(\hat p_0)\\
&=p_0\tilde D_x(\underbrace{r_{1,z}+u_xu_z-u_t}_{=\, 0 \ \mathrm{due\ to\ Eq.}\eqref{yme-taur}})
+u_x(\underbrace{\tilde D_z(p_r^1)-\tilde D_t(p_0)+u_x\tilde D_z(p_0)+u_z\tilde D_x(p_0)}_{=\, 0 \ \mathrm{due\ to\ Eq.}\eqref{nsym-yme7}})\\
&-(\underbrace{\tilde D_z(p_2^r)-u_z\tilde D_x(p_1^r)-r_{1,x}\tilde D_z(p_0)+\tilde D_t(p_1^r)}_{=\, 0 \ \mathrm{due\ to\ Eq.}\eqref{nsym-yme9}}) = 0,
\end{align*}
thus $\hat p_1^r$ satisfies the conditions \eqref{nsym-yme6} and \eqref{nsym-yme7} on the equation $\tilde{\mathcal{E}}^W$.\\

\item[5)] For each $\beta \geq 2$, we have:
\begin{align*}
\tilde D_x(\hat p_{\beta}^r)&-u_x\tilde D_x(\hat p_{\beta-1}^r)-r_{\beta-1,x}\tilde D_x(\hat p_0)+\tilde D_y(\hat p_{\beta-1}^r)\\
&=p_0\tilde D_x(\underbrace{r_{\beta,x}-u_xr_{\beta-1,x}+r_{\beta-1,y}}_{=\, 0 \ \mathrm{due\ to\ Eq.}\eqref{yme-taur}}) - r_{\beta-1,x}(\underbrace{\tilde D_x(p_1^r)-\tilde D_y(p_0)+2u_x\tilde D_x(p_0)}_{=\, 0 \ \mathrm{due\ to\ Eq.}\eqref{nsym-yme6}})\\
&-(\underbrace{\tilde D_x(p_{\beta+1}^r)-u_x\tilde D_x( p_{\beta}^r)-r_{\beta,x}\tilde D_x(p_0)+\tilde D_y(p_{\beta}^r)}_{=\, 0 \ \mathrm{due\ to\ Eq.}\eqref{nsym-yme8}})=0,
\intertext{and also}
\tilde D_z(\hat p_{\beta}^r)&-u_z\tilde D_x(\hat p_{\beta-1}^r)-r_{\beta-1,x}\tilde D_z(\hat p_0)+\tilde D_t(\hat p_{\beta-1}^r)\\
&=p_0\tilde D_x (\underbrace{r_{\beta,z}-u_zr_{\beta-1,x}+r_{\beta-1,t}}_{=\, 0 \ \mathrm{due\ to\ Eq.}\eqref{yme-taur}})\\
&-r_{\beta-1,x}(\underbrace{\tilde D_z(p_1^r)-\tilde D_t(p_0)+u_x\tilde D_z(p_0)+u_z\tilde D_x(p_0)}_{=\, 0 \ \mathrm{due\ to\ Eq.}\eqref{nsym-yme7}})\\
&-(\underbrace{\tilde D_z(p_{\beta+1}^r)-u_z\tilde D_x( p_{\beta}^r)-r_{\beta,x}\tilde D_z(p_0)+\tilde D_t(p_{\beta}^r)}_{=\, 0 \ \mathrm{due\ to\ Eq.}\eqref{nsym-yme9}}) =0,
\end{align*}
thus $\hat p_\beta^r$ satisfies the conditions \eqref{nsym-yme8} and \eqref{nsym-yme9} on the equation $\tilde{\mathcal{E}}^W$.
\end{itemize}
\qed

\section*{Appendix B: Proof of Proposition \ref{ROm}}
Let $P \in {\mathrm{sym}}_{\mathcal{L}}^{\tau^W}(\mathcal{E})$  be an arbitrary $\tau^W$-symmetry of the rYME \eqref{yme}, and denote $\hat P = \mathcal R^q(P)$, $\bar P = \mathcal R^m(P)$. Then $\hat P \in {\mathrm{sym}}_{\mathcal{L}}^{\tau^W}(\mathcal{E})$ and, according to \eqref{yme-ro2}, it holds 
$$\bar P=\mathcal{R}^m (P)=(z\mathcal{I}+t\mathcal{R}^q)(P)=t \hat P + zP.$$
We need to show that $\bar P \in {\mathrm{sym}}_{\mathcal{L}}^{\tau^W}(\mathcal{E})$.

For brevity, let us denote by $L_i$, $i=1,\ldots,9$, the left-hand sides of the respective conditions \eqref{nsym-yme1}-\eqref{nsym-yme9}. Then $L_i(P)=L_i(\hat P)=0$ on $\tilde{\mathcal E}^W$ ($L_i(P)$ means the substitution of the corresponding components of $P$ to $L_i$, etc.), and one can obtain by straightforward computations for each $\alpha\geq 0,\beta\geq 2$ the following:
\begin{align*}
L_1(\bar P)&=tL_1(\hat P)+zL_1(P)-\tilde{D}_x(\hat p_0)+\tilde{D}_y(p_0)-u_x\tilde{D}_x(p_0)+u_{xx} p_0\\[2mm]
&=-\tilde D_x(p_1^r)+\tilde{D}_y(p_0)-2u_x\tilde{D}_x(p_0) = -L_6(P)=0,\\[2mm]
L_2(\bar P)&=tL_2(\hat P) + zL_2(P) + \underbrace{\hat p_{\alpha}^q-q_{\alpha,x}p_0-p_{\alpha-1}^q}_{=0 \ \mathrm{due\ to\ Eq.} \eqref{yme-ro1-2}}  = 0,\\[2mm]
L_3(\bar P)&=tL_3(\hat P) + zL_3(P) = 0\\[2mm]
L_4(\bar P)&=tL_4(\hat P) + zL_4(P) +\underbrace{\hat p_{\alpha}^m-m_{\alpha,x}p_0+\frac ztp_{\alpha}^m+\frac 1t p_{\alpha-1}^m}_{=0 \ \mathrm{due\ to\ Eq.} \eqref{yme-ro1-3}} = 0\\[2mm]
L_5(\bar P)&=tL_5(\hat P) + zL_5(P) = 0\\[2mm]
L_6(\bar P)&=tL_6(\hat P) + zL_6(P) = 0\\[2mm]  
L_7(\bar P)&=tL_7(\hat P) + zL_7(P) - \underbrace{\hat p_0+p_1^r+u_xp_0}_{=0 \ \mathrm{due\ to\ Eq.} \eqref{yme-ro1-1}}=0\\[2mm]
L_8(\bar P)&=tL_8(\hat P) + zL_8(P) = 0\\[2mm] 
L_9(\bar P)&=tL_9(\hat P) + zL_9(P) + \underbrace{\hat p_{\beta-1}^r+p_{\beta}^r-r_{{\beta}-1,x}p_0}_{=0 \ \mathrm{due\ to\ Eq.} \eqref{yme-ro1-4}}= 0.
\end{align*}
where $p_{-1}^q=p_{-1}^m=0$. Thus, the components od $\bar P$ satisfy all conditions \eqref{nsym-yme1}-\eqref{nsym-yme9} on $\tilde{\mathcal E}^W$ which implies $\bar P \in {\mathrm{sym}}_{\mathcal{L}}^{\tau^W}(\mathcal{E})$.
\qed

\section*{Appendix C: Left-upper corners of the matrix representations of the recursion operators $({\mathcal R}^q)^{-1}$, $\mathcal R^m$,  and $({\mathcal R}^m)^{-1}$}
\begin{equation*}
({\mathcal R}^q)^{-1}=
\begin{pmatrix}
0 & 1/q_{0,x} & 0 & 0 & 0 & 0 & 0 & 0 & 0 & 0 & 0 & 0 & \ldots\\[2mm]
0 & -q_{1,x}/q_{0,x} & 0 & 0 & 1 & 0 & 0 & 0 & 0 & 0 & 0 & 0 & \ldots\\[2mm]
0 & -\frac{t}{q_{0,x}} \cdot \sum \limits_{j=0}^0 \frac{m_{j,x}}{(-z)^{1-j}} & -\frac{t}{z} & 0 & 0 & 0 & 0 & 0 & 0 & 0 & 0 & 0 & \ldots\\[2mm]
1 & r_{0,x}/q_{0,x} & 0 & 0 & 0 & 0 & 0 & 0 & 0 & 0 & 0 & 0 & \ldots\\[2mm]
0 & -q_{2,x}/q_{0,x} & 0 & 0 & 0 & 0 & 0 & 1 & 0 & 0 & 0 & 0 & \ldots\\[2mm]
0 & -\frac{t}{q_{0,x}} \cdot \sum \limits_{j=0}^1 \frac{m_{j,x}}{(-z)^{2-j}} & \frac{t}{z^2} & 0 & 0 & -\frac{t}{z} & 0 & 0 & 0 & 0 & 0 & 0 & \ldots\\[2mm]
0 & r_{1,x}/q_{0,x} & 0 & -1 & 0 & 0 & 0 & 0 & 0 & 0 & 0 & 0 & \ldots\\[2mm]
0 & -q_{3,x}/q_{0,x} & 0 & 0 & 0 & 0 & 0 & 0 & 0 & 0 & 1 & 0 & \ldots\\[2mm]
0 & -\frac{t}{q_{0,x}} \cdot \sum \limits_{j=0}^2 \frac{m_{j,x}}{(-z)^{3-j}} & -\frac{t}{z^3} & 0 & 0 & \frac{t}{z^2} & 0 & 0 & -\frac{t}{z} & 0 & 0 & 0 & \ldots\\[2mm]
0 & r_{2,x}/q_{0,x} & 0 & 0 & 0 & 0 & -1 & 0 & 0 & 0 & 0 & 0 & \ldots\\[2mm]
\vdots & \vdots & \vdots & \vdots & \vdots & \vdots & \vdots & \vdots & \vdots & \vdots & \vdots & \vdots & \ddots
\end{pmatrix}\\[5mm]
\end{equation*}

\begin{equation*}
{\mathcal R^m}=
\begin{pmatrix}
z+t u_x & 0 & 0 & t & 0 & 0 & 0 & 0 & 0 & 0 & 0 & 0 & 0 & 0 & \ldots\\[2mm]
tq_{0,x} & z & 0 & 0 & 0 & 0 & 0 & 0 & 0 & 0 & 0 & 0 & 0 & 0 & \ldots\\[2mm]
tm_{0,x} & 0 & 0 & 0 & 0 & 0 & 0 & 0 & 0 & 0 & 0 & 0 & 0 & 0 & \ldots\\[2mm]
tr_{1,x} & 0 & 0 & z & 0 & 0 & -t & 0 & 0 & 0 & 0 & 0 & 0 & 0 & \ldots\\[2mm]
tq_{1,x} & t & 0 & 0 & z & 0 & 0 & 0 & 0 & 0 & 0 & 0 & 0 & 0 & \ldots\\[2mm]
tm_{1,x} & 0 & -1 & 0 & 0 & 0 & 0 & 0 & 0 & 0 & 0 & 0 & 0 & 0 & \ldots\\[2mm]
tr_{2,x} & 0 & 0 & 0 & 0 & 0 & z & 0 & 0 & -t & 0 & 0 & 0 & 0 & \ldots\\[2mm]
tq_{2,x} & 0 & 0 & 0 & t & 0 & 0 & z & 0 & 0 & 0 & 0 & 0 & 0 & \ldots\\[2mm]
tm_{2,x} & 0 & 0 & 0 & 0 & -1& 0 & 0 & 0 & 0 & 0 & 0 & 0 & 0 & \ldots\\[2mm]
tr_{3,x} & 0 & 0 & 0 & 0 & 0 & 0 & 0 & 0 & z & 0 & 0 & -t & 0 & \ldots \\[2mm]
\vdots & \vdots & \vdots & \vdots & \vdots & \vdots & \vdots & \vdots & \vdots & \vdots & \vdots & \vdots & \vdots & \vdots & \ddots
\end{pmatrix}
\end{equation*}

\begin{equation*}
({\mathcal R}^m)^{-1}=
\begin{pmatrix}
0 & 0 & 1/(tm_{0,x}) & 0 & 0 & 0 & 0 & 0 & 0 & \ldots\\[2mm]
0 & \frac{1}{z} & - \frac{1}{m_{0,x}}\sum \limits_{j=0}^{0}\frac{(-t)^{-j}q_{j,x}}{z^{1-j}} & 0 & 0 & 0 & 0 & 0 & 0 & \ldots\\[2mm]
0 & 0 & m_{1,x}/m_{0,x} & 0 & 0 & -1 & 0 & 0 & 0 &  \ldots\\[2mm]
\frac{1}{t} & 0 &  \frac{1}{m_{0,x}}\left(-\frac{z}{t^2}+\sum \limits_{j=1}^{1}\frac{z^{1-j}r_{j-1,x}}{t^{2-j}}\right) & 0 & 0 & 0 & 0 & 0 & 0 &  \ldots\\[2mm]
0 & -\frac{t}{z^2} & - \frac{1}{m_{0,x}}\sum \limits_{j=0}^{1}\frac{(-t)^{1-j}q_{j,x}}{z^{2-j}} & 0 & \frac{1}{z} & 0 & 0 & 0 & 0 & \ldots\\[2mm]
0 & 0 & m_{2,x}/m_{0,x} & 0 & 0 & 0 & 0 & 0 & -1 & \ldots\\[2mm]
\frac{z}{t^2} & 0 &  \frac{1}{m_{0,x}}\left(-\frac{z^2}{t^3}+\sum \limits_{j=1}^{2}\frac{z^{2-j}r_{j-1,x}}{t^{3-j}}\right) & -\frac{1}{t} & 0 & 0 & 0 & 0 & 0 & \ldots\\[2mm]
0 & \frac{t^2}{z^3} &- \frac{1}{m_{0,x}}\sum \limits_{j=0}^{2}\frac{(-t)^{2-j}q_{j,x}}{z^{3-j}} & 0 & -\frac{t}{z^2} & 0 & 0 & \frac{1}{z} & 0 & \ldots\\[2mm]
0 & 0 & m_{3,x}/m_{0,x} & 0 & 0 & 0 & 0 & 0 & 0 & \ldots\\[2mm]
\frac{z^2}{t^3} & 0 & \frac{1}{m_{0,x}}\left(-\frac{z^3}{t^4}+\sum \limits_{j=1}^{3}\frac{z^{3-j}r_{j-1,x}}{t^{4-j}}\right) & -\frac{z}{t^2} & 0 & 0 & -\frac{1}{t} & 0 & 0 & \ldots \\[2mm]
\vdots & \vdots & \vdots & \vdots & \vdots & \vdots & \vdots & \vdots & \vdots & \ddots
\end{pmatrix}
\end{equation*}


\begin{thebibliography}{99}
\bibitem{Ablowitz91} M. J. Ablowitz, P. A. Clarkson. \emph{Solitons, nonlinear evolution equations and inverse scattering}. Cambridge University Press, Cambridge, 1991. \href{https://doi.org/10.1017/CBO9780511623998}{DOI:10.1017/CBO9780511623998}.
\bibitem{Bar} H.~Baran, \emph{Infinitely many commuting nonlocal symmetries for modified Mart\'inez Alonso-Shabat equation}, Commun. Nonlinear Sci. Numer. Simul. \textbf{96} (2021), Paper No. 105692, 4 pp. \href{https://doi.org/10.1016/j.cnsns.2021.105692}{DOI:10.1016/j.cnsns.2021.105692}, \href{https://arxiv.org/abs/1911.08985}{arXiv:1911.08985}
\bibitem{BKMV} H.~Baran, I.S.~Krasil{\cprime}shchik, O.I.~Morozov, and P.~Voj\v{c}\'{a}k, \emph{Nonlocal Symmetries of Integrable Linearly Degenerate Equations: A Comparative Study}, Theor.\ and Math.\ Phys., \textbf{196} (2018), Issue 2, 1089--1110. \href{https://doi.org/10.1134/S0040577918080019}{DOI:10.1134/S0040577918080019}, \href{https://arxiv.org/abs/1611.04938}{arXiv:1611.04938}.
\bibitem{Jets} H.~Baran, M.~Marvan, \emph{Jets. A software for differential calculus on jet spaces and diffieties}.  \url{http://jets.math.slu.cz}.
\bibitem{Bluman89} G. W. Bluman, S. Kumei. \emph{Symmetries and Differential Equations}. Springer-Verlag, 1989.
\bibitem{boch} A.V. Bocharov et al., \emph{Symmetries and Conservation Laws for Differential Equations of Mathematical Physics}, AMS, Providence, RI, 1999.
\bibitem{bog} L.V. Bogdanov, M.V. Pavlov. \emph{Linearly degenerate hierarchies of quasiclassical SDYM type}. J. Math. Phys. 58 (2017), 093505. \href{https://doi.org/10.1063/1.5004258}{DOI:10.1063/1.5004258}, \href{https://arxiv.org/abs/1603.00238v2}{arXiv:1603.00238v2}.
\bibitem{DFKN35} B. Doubrov, E.V. Ferapontov, B. Kruglikov, V.S. Novikov, \emph{On integrability in Grassmann geometries: integrable systems associated with fourfolds in \textbf{Gr}(3,5)}.  Proceedings of the London Mathematical Society 116.5 (2018), 1269--1300. \href{https://doi.org/10.1112/plms.12114}{DOI:10.1112/plms.12114}, \href{https://arxiv.org/abs/1503.02274}{ arXiv:1503.02274v2}
\bibitem{DFKN} B.~Doubrov, E.~Ferapontov, B.~Kruglikov, V.~Novikov, \emph{Integrable systems in 4D associated with sixfolds in $\mathbf{Gr}(4,6)$}, International Math. Research Notices \textbf{21} (2019), 6585--6613. \href{https://doi.org/10.1093/imrn/rnx308}{doi:10.1093/imrn/rnx308}, \href{https://arxiv.org/abs/1705.06999}{arXiv:1705.06999}
\bibitem{Dunajski10} M. Dunajski. \emph{Solitons, Instantons and Twistors}. Oxford University Press, 2010.
\bibitem{Dun} M.~Dunajski, W.~Kry\'{n}ski, \emph{Einstein-Weyl geometry, dispersionless Hirota equation and Veronese webs}, Math.\ Proc.\ Cambridge Philosophical Society \textbf{157} (2014) 139--150, \href{https://doi.org/10.1017/S0305004114000164}{DOI:10.1017/S0305004114000164}, \href{https://arxiv.org/abs/1301.0621}{arXiv:1301.0621}.
\bibitem{fer} E.V. Ferapontov, K.R. Khusnutdinova, \emph{Hydrodynamic reductions of multi-dimensional dispersionless PDEs: the test for integrability}. J. Math. Phys. \textbf{45} (2004), 2365--2377. \href{https://doi.org/10.1063/1.1738951}{DOI:10.1063/1.1738951},  \href{https://arxiv.org/abs/nlin/0312015}{arXiv:nlin/0312015v1}
\bibitem{FerMoss} E.V. Ferapontov, J. Moss,\emph{ Linearly degenerate partial differential equations and quadratic line complexes}, Commun. Anal. Geom., 23, 91--127 (2015). \href{https://doi.org/10.4310/CAG.2015.v23.n1.a3}{DOI:10.4310/CAG.2015.v23.n1.a3},  \href{https://arxiv.org/abs/1204.2777v1}{arXiv:1204.2777v1}.
\bibitem{Fokas87} A. S. Fokas, \emph{Symmetries and integrability.} Stud. Appl. Math. 77 (1987) 253--299.
\bibitem{kramor} I.S. Krasil'shchik, O.I. Morozov, \emph{Lagrangian extensions of multi-dimensional integrable equations. I. The five-dimensional Mart\'inez Alonso-Shabat equation}. Anal. Math. Phys. \textbf{13} (2023), no. 1, 20 pp. \href{https://doi.org/10.1007/s13324-022-00763-w}{DOI:10.1007/s13324-022-00763-w}, \href{https://arxiv.org/abs/2207.07936}{ arXiv:2207.07936v1}
\bibitem{KMV} I.S. Krasil'shchik, O.I. Morozov, P. Voj\v c\'{a}k, \emph{Nonlocal symmetries, conservation laws, and recursion operators of the Veronese web equation}, J. Geom. Phys. 146 (2019),103519. \href{https://doi.org/10.1016/j.geomphys.2019.103519}{DOI:10.1016/j.geomphys.2019.103519}, \href{https://arxiv.org/abs/1902.09341}{arXiv:1902.09341v3}.
\bibitem{kra1} I.S. Krasil{\cprime}shchik, A. M. Verbovetsky,  \emph{Geometry of jet spaces and integrable systems}, J. Geom. Phys. {\bf 61} (2011), no. 9, 1633--1674. \href{https://doi.org/10.1016/j.geomphys.2010.10.012}{DOI:10.1016/j.geomphys.2010.10.012}, \href{https://arxiv.org/abs/1002.0077}{arXiv:1002.0077v6}
\bibitem{kra2} I.S. Krasil{\cprime}shchik, A.M. Verbovetsky, R. Vitolo, \emph{The Symbolic Computation of Integrability Structures for Partial Differential Equations}, Texts \& Monographs in Symbolic Computation, Springer, Berlin (2017). \href{https://doi.org/10.1007/978-3-319-71655-8}{DOI:10.1007/978-3-319-71655-8}.
\bibitem{kra3} I.S. Krasil{\cprime}shchik, A. M. Vinogradov,  \emph{Nonlocal trends in the geometry of differential equations: symmetries, conservation laws, and B\"{a}cklund transformations. Symmetries of partial differential equations}, Part I. Acta Appl. Math.  {\bf 15}  (1989),  no. 1-2, 161--209. \href{https://doi.org/10.1007/BF00131935}{DOI:10.1007/BF00131935}
\bibitem{KrasVoj} Krasil'shchik I.S., Voj\v{c}\'{a}k P., \emph{On the algebra of nonlocal symmetries for the \textrm{4D} Mart\'{i}nez Alonso-Shabat equation}. J. Geom. Phys. 2011;61(9):1633-74. \href{https://doi.org/10.1016/j.geomphys.2021.104122}{DOI:10.1016/j.geomphys.2021.104122}, \href{https://arxiv.org/pdf/2008.10281.pdf}{arXiv:2008.10281v1}
\bibitem{KrugMor2016} B.S. Kruglikov, O.I. Morozov, \emph{A B\"{a}cklund transformation between the four-dimensional Mart\'{i}nez Alonso-Shabat and Ferapontov-Khusnutdinova equations}, Theor. Math. Phys. 188 (3) (2016) 1358-1360. \href{https://doi.org/10.1134/S0040577916090063}{DOI:10.1134/S0040577916090063}, \href{https://arxiv.org/abs/1502.00902}{arXiv:1502.00902v1}
\bibitem{Kru} B.~Kruglikov, A.~Panasyuk, \emph{Veronese webs and nonlinear PDEs}, J.\ of Geom.\ and Phys.\ \textbf{115} (2017), 45--60, \href{https://doi.org/10.1016/j.geomphys.2016.08.008}{DOI:10.1016/j.geomphys.2016.08.008}, \href{https://arxiv.org/abs/1602.07346}{arXiv:1602.07346}.
\bibitem{Man} S.V. Manakov, P.M. Santini, \emph{A hierarchy of integrable partial differential equations in 2+1 dimensions associated with one-parameter families of one-dimensional vector fields}. Theor. Math. Phys., {\bf 152} (2007), 1004--1011.  \href{https://doi.org/10.4213/tmf6076}{DOI:10.4213/tmf6076}, \href{https://arxiv.org/abs/nlin/0611047}{arXiv:nlin/0611047}
\bibitem{Marvan95} M. Marvan, \emph{Another Look on Recursion Operators}, in \emph{Differential Geometry and Applications}, Proc. Conf. 393, Brno, 1995.
\bibitem{mor} O.I. Morozov, \emph{Isospectral deformation of the reduced quasi-classical self-dual Yang-Mills equation}. Differential Geom. Appl. \textbf{76} (2021), 101742, 14 pp. \href{https://doi.org/10.1016/j.difgeo.2021.101742}{DOI:10.1016/j.difgeo.2021.101742}, \href{https://arxiv.org/abs/2012.06904}{arXiv:2012.06904}
\bibitem{Mor-Ser} O.I.~Morozov, A.~Sergyeyev, \emph{The four-dimensional Mart\'{\i}nez Alonso-Shabat equation: reductions and nonlocal symmetries}. J.\ of Geom.\ and Phys. \textbf{85} (2014), 40--45. \href{https://doi.org/10.1016/j.geomphys.2014.05.025}{DOI:10.1016/j.geomphys.2014.05.025},  \href{https://arxiv.org/abs/1401.7942}{arXiv:1401.7942v2}
\bibitem{Olver77} P. J. Olver. \emph{Evolution equations possessing infinitely many symmetries}. Journal of Mathematical Physics, 18:1212--1215, 1977. \href{https://doi.org/10.1063/1.523393}{DOI:10.1063/1.523393}
\bibitem{Olver93} P. J. Olver. \textit{Applications of Lie Groups to Differential Equations}, 2nd edition, Springer-Verlag, 1993.
\bibitem{Papachristou91} C. J. Papachristou. \emph{Lax pair, hidden symmetries, and infinite sequences of conserved currents for self-dual Yang-Mills fields}. J. Phys. A: Math. Gen. 24 (1991) L1051--L1055. \href{https://doi.org/10.1088/0305-4470/24/17/015}{DOI 10.1088/0305-4470/24/17/015}
\bibitem{Pav} M.V. Pavlov, \emph{Integrable hydrodynamic chains}, J. Math. Phys. {\bf 44} (2003), 4134--4156. \href{https://doi.org/10.1063/1.1597946}{DOI:10.1063/1.1597946}, \href{https://arxiv.org/abs/nlin/0301010}{ arXiv:nlin/0301010v1}
\bibitem{Serg} A. Sergyeyev, \emph{A simple construction of recursion operators for multidimensional dispersionless integrable systems}. J. Math. Anal. Appl., \textbf{454} (2017), 468--80. \href{https://doi.org/10.1016/j.jmaa.2017.04.050}{DOI:10.1016/j.jmaa.2017.04.050},  \href{https://arxiv.org/abs/1501.01955}{arXiv: 1501.01955}
\bibitem{voj} P. Voj\v c\'ak, \emph{Non-Abelian covering and new recursion operators for the 4D Mart\'inez Alonso-Shabat equation}. Commun. Nonlinear Sci. Numer. Simul. {\bf 118} (2023), Paper No. 107007, 11 pp. \href{https://doi.org/10.1016/j.cnsns.2022.107007}{DOI:10.1016/j.cnsns.2022.107007}, \href{https://arxiv.org/abs/2206.10530}{ arXiv:2206.10530v1}
\bibitem{Wahlquist} H.D. Wahlquist, F. B. Estabrook. \emph{Prolongation structures of nonlinear evolution equations}, J. Math. Phys. 16 (1975), 1--7. \href{https://doi.org/10.1063/1.522396}{DOI:10.1063/1.522396}
\bibitem{Zakh} I.~Zakharevich, \emph{Nonlinear wave equation, nonlinear Riemann problem, and the twistor transform of Veronese webs}. \href{https://arxiv.org/abs/math-ph/0006001}{arXiv:math-ph/0006001}

\end{thebibliography}
\end{document}